%% file: cvpr25_text2diagram.tex
\documentclass[10pt,twocolumn,letterpaper]{article}

\usepackage{cvpr}              % To produce the CAMERA-READY version
\usepackage{booktabs}
\usepackage{amsmath}
\usepackage{multirow}
\usepackage{listings}
\usepackage{algorithm}
\usepackage{algorithmic}
\usepackage{pifont}


\definecolor{cvprblue}{rgb}{0.21,0.49,0.74}
\usepackage[pagebackref,breaklinks,colorlinks,allcolors=cvprblue]{hyperref}

\def\paperID{12968} % *** Enter the Paper ID here
\def\confName{CVPR}
\def\confYear{2025}

\vspace{-2mm}
\title{From Words to Structured Visuals: A Benchmark and Framework for Text-to-Diagram Generation and Editing}

\author{
    Jingxuan Wei$^{1*}$, Cheng Tan$^{2*}$, Qi Chen$^{1*}$, Gaowei Wu$^{1}$\thanks{Equal contribution}, Siyuan Li$^{2}$, Zhangyang Gao$^{2}$, \\
    Linzhuang Sun$^{1}$, Bihui Yu$^{1}$, Ruifeng Guo$^{1}$  \\
    $^1$University of Chinese Academy of Sciences 
    $^2$ Westlake University \\
    {\tt\small weijingxuan20@mails.ucas.edu.cn, tancheng@westlake.edu.cn}
}

\begin{document}
\maketitle

\begin{abstract}
We introduce the task of text-to-diagram generation, which focuses on creating structured visual representations directly from textual descriptions. Existing approaches in text-to-image and text-to-code generation lack the logical organization and flexibility needed to produce accurate, editable diagrams, often resulting in outputs that are either unstructured or difficult to modify. To address this gap, we introduce DiagramGenBenchmark, a comprehensive evaluation framework encompassing eight distinct diagram categories, including flowcharts, model architecture diagrams, and mind maps. Additionally, we present DiagramAgent, an innovative framework with four core modules—Plan Agent, Code Agent, Check Agent, and Diagram-to-Code Agent—designed to facilitate both the generation and refinement of complex diagrams. Our extensive experiments, which combine objective metrics with human evaluations, demonstrate that DiagramAgent significantly outperforms existing baseline models in terms of accuracy, structural coherence, and modifiability. This work not only establishes a foundational benchmark for the text-to-diagram generation task but also introduces a powerful toolset to advance research and applications in this emerging area.
\end{abstract}

\section{Introduction}
\textit{``Drawing is not what one sees but what one can make others see." — Edgar Degas.} 

The task of transforming textual descriptions into structured diagrams, referred to as \textit{text-to-diagram generation}, is crucial for domains that rely on clear, logical representations of complex information~\cite{mccracken2001text,guo2024survey,cao2024survey}. Unlike natural scene images, structured diagrams—such as flowcharts, mind maps, and model architecture diagrams—require precise relationships and logical coherence to effectively communicate intricate ideas. This study focuses on developing robust methods for automatic diagram generation and editing, aiming to facilitate efficient, accurate diagram creation from textual input. Such advancements hold great potential to enhance productivity in domains like education, scientific research, and industry, where structured visualizations are essential for effective communication and analysis.

While recent advancements in text-to-image generation have produced impressive results in synthesizing photorealistic and aesthetically pleasing images~\cite{zhou2023vision+, cao2024survey}, these methods are largely optimized for naturalistic visuals rather than structured, logical content. Models such as GANs~\cite{stackgan, attngan} and diffusion models~\cite{photorealistic, precisecontrol} have excelled at generating images with rich visual detail, but they inherently lack the ability to enforce the precise relationships and hierarchical organization required for structured diagrams. Alternative approaches have sought to first convert text into code representations~\cite{allamanis2018survey,kumar2023comprehensive}, which are then compiled into charts or simple plots. These methods leverage advancements in large language models to interpret textual descriptions as code that can produce visual output. However, while effective for generating basic visualizations such as bar charts and line graphs, this approach has limitations when applied to more complex diagrams. Specifically, they lack the flexibility and structural logic required for diagrams that demand intricate organization and interactivity to edits—a necessity for broader scientific and industrial applications.

\begin{figure*}[ht]
    \centering
    \includegraphics[width=\textwidth]{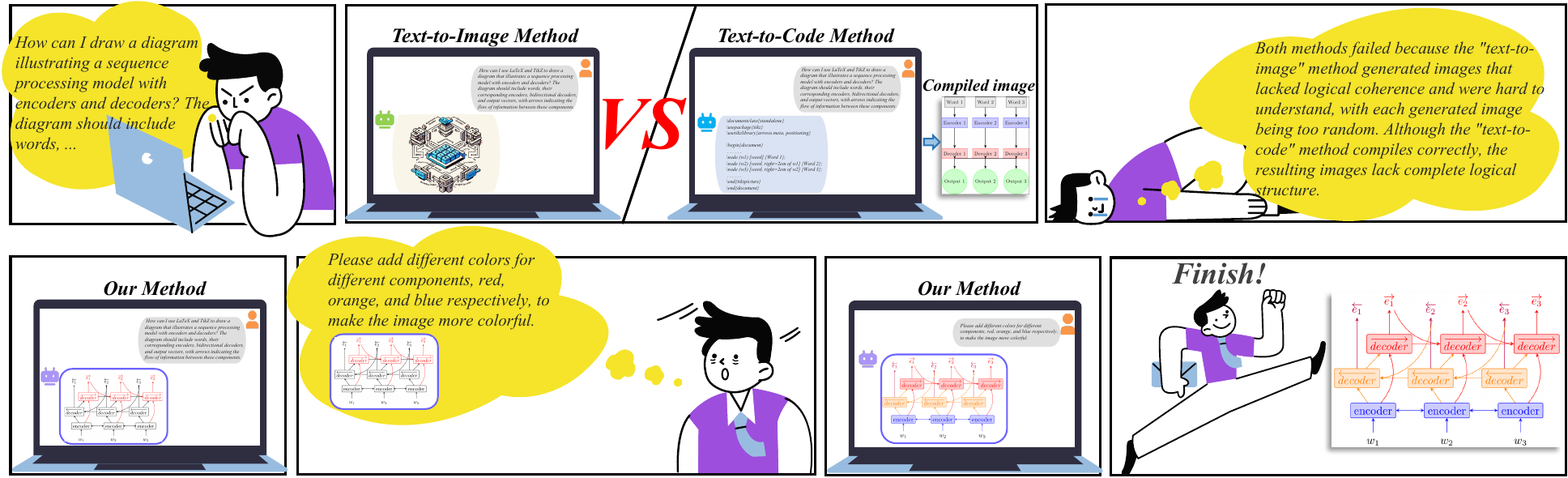}
    \caption{Challenges in Existing Text-to-Image and Text-to-Code Methods for Diagram Generation}
    \label{fig:challenges}
\end{figure*}

As illustrated in Figure~\ref{fig:challenges}, existing methods fall short when faced with tasks that require both logical precision and post-generation modifiability. Text-to-image models often produce outputs that are visually appealing but logically incoherent~\cite{lin2025evaluating, zhou2023vision+}, failing to capture the structured relationships that intricate diagrams demand. This gap in logical structure limits the practical utility of these models in scenarios where accurate and adaptable visual representations are essential. On the other hand, current text-to-code approaches, while suitable for simpler visualizations, struggle to represent complex diagram features such as hierarchical relationships, intricate color coding, and layered structures~\cite{allamanis2018survey, kumar2023comprehensive}. Furthermore, there is a significant lack of comprehensive datasets tailored for generating and modifying diverse types of structured diagrams. Existing datasets tend to focus on either image realism or simplistic data visualizations, neither of which fully address the demands of text-to-diagram generation. 
% This data gap underscores the need for a specialized benchmark that can accommodate a variety of structured diagram types and support both their generation and iterative refinement.   %需要再次寻思一下，直接注释掉？

To address these challenges, we introduce \textbf{DiagramGenBenchmark}, a rigorously curated dataset for evaluating structured diagram generation and editing. This benchmark includes eight distinct types of diagrams, such as flowcharts, bar charts, and model architectures, providing a solid foundation for advancing and assessing text-to-diagram tasks. Alongside this benchmark dataset, we propose \textbf{DiagramAgent}, a framework for text-to-diagram generation and editing. DiagramAgent integrates multiple agents, including a \textit{Plan Agent} for interpreting textual commands, a \textit{Code Agent} for generating diagram code, a \textit{Check Agent} for logical verification, and a \textit{Diagram-to-Code Agent}. This framework not only facilitates robust diagram generation but also enables editing and adaptation, accommodating complex structural requirements.

Our extensive experiments on DiagramAgent, conducted using the DiagramGenBenchmark dataset, demonstrate its superior performance across various diagram types and tasks. Notably, DiagramAgent outperforms existing works in terms of accuracy and overall quality. For example, in the diagram generation task, DiagramAgent achieved top scores with Pass@1 reaching 58.15, ROUGE-L scoring 51.97, and CodeBLEU at 86.83, highlighting its efficacy in generating accurate and logically coherent diagram code. We further validated DiagramAgent's versatility by testing it on both open-source models, such as Qwen2~\cite{qwen2}, and closed-source models, including GPT-4o~\cite{gpt4}, demonstrating its adaptability across different model architectures. To substantiate these results, we conducted human evaluations that aligned closely with the objective metrics, confirming DiagramAgent's accuracy and usability in diagram generation tasks. We summarize our contributions as follows:
\begin{itemize}
    \item We introduce \textbf{DiagramGenBenchmark}, a benchmark dataset covering eight diverse types of diagrams, specifically tailored for diagram generation and editing.
    \item We present \textbf{DiagramAgent}, a text-to-diagram framework capable of both generating and editing diagrams, providing flexibility and accuracy in complex logical structures.
    \item We conduct comprehensive experiments demonstrating DiagramAgent's superior performance in terms of accuracy and logical coherence, establishing DiagramAgent as a reliable baseline for future research.
\end{itemize}

\section{Related Work}

\subsection{Text-to-image generation}
Text-to-image generation involves creating images based on natural language descriptions, ranging from photorealistic scenes to semantically rich and abstract visuals~\cite{zhou2023vision+,cao2024survey}. Existing methods can be broadly categorized into three main approaches: Generative Adversarial Networks (GANs), diffusion models, and autoregressive approaches. GANs represent some of the earliest efforts in this field, with models such as StackGAN~\cite{stackgan} and AttnGAN~\cite{attngan} improving image quality through progressive generation strategies and attention mechanisms. Despite their early successes, GANs are often hampered by issues like instability in adversarial training and mode collapse, making them less suitable for generating images of complex scenes. Diffusion models have emerged as the dominant approach, offering significant improvements in image quality by iteratively refining images through a denoising process. Notable examples include Imagen~\cite{photorealistic}, which generates high-resolution images with remarkable detail. PreciseControl~\cite{precisecontrol} enable fine-grained control over specific attributes, while OpenBias~\cite{openbias} incorporates mechanisms for bias detection to ensure fairness and diversity in generated outputs. Transformer-based autoregressive models have also gained prominence due to their strong cross-modal alignment capabilities. For instance, DALL-E~\cite{ramesh2021zero} employs attention mechanisms to closely map textual descriptions to visual elements. The STAR model~\cite{star} integrates multi-scale autoregressive generation with enhanced Rotary Position Embeddings (RoPE) to improve consistency between text and images in intricate scenarios. Similarly, Fluid~\cite{fluid} leverages multi-level attention to fine-tune the alignment of object layouts and sizes, ensuring a precise correspondence between textual inputs and visual outputs. While these methods have made significant strides in creating natural scene imagery, they fall short when it comes to tasks that require the generation of diagrams with logical structures and explicit relationships, such as flowcharts, architecture diagrams, or other technical illustrations.
%  In contrast, our proposed method addresses these limitations by not only generating accurate diagrammatic representations but also supporting multiple edits and modifications, thereby enabling an interactive and adaptable text-to-diagram generation process.

\subsection{Text-to-code generation}

Text-to-code generation involves the automatic conversion of natural language descriptions into executable code. Notable examples include Qwen2.5-Coder~\cite{qwen25}, DeepSeek-Coder-Instruct~\cite{deepseek}, Code-Llama~\cite{codellama}, and WizardCoder~\cite{wizardcoder}, which excel at generating sophisticated code snippets that integrate architectural logic. In contrast, generalist text-to-code models, although not explicitly designed for code generation, have demonstrated remarkable adaptability due to their extensive pre-trainings. Open-sourced models such as Llama3~\cite{llama3}, Baichuan2~\cite{baichuan2}, InternLM2.5~\cite{internlm2}, and Qwen2~\cite{qwen2} exhibit strong capabilities across multiple programming languages, making them well-suited for a broad range of coding challenges. Moreover, several closed-sourced models, including GPT-4o~\cite{gpt4}, Gemini~\cite{gemini}, DeepSeek-2.5~\cite{deepseekv2}, and GLM-4-Plus~\cite{chatglm}, offer exceptional accuracy and performance in complex code generation. 
% However, their closed-source nature restricts user customization and limits transparency regarding their underlying mechanisms despite their performance advantages. 
Text-to-code models are also paving the way for innovations that extend beyond conventional programming tasks. The methodologies employed in text-to-code generation have inspired research into adjacent areas, such as automatic diagram generation in this work.

% \subsection{Image-to-code generation}
% The image-to-code generation task involves the automatic conversion of the diagram image inputs into code~\cite{unifying,AutomaTikZ}. Specialized models target specific tasks, primarily focusing on chart-to-code and formula-to-code conversions. The chart-to-code models, such as Plot2Code~\cite{plot2code} and ChartMimic~\cite{chartmimic}, aim to translate visual representations of charts (e.g., line charts, bar charts) into corresponding code. Similarly, formula-to-code models, such as MathLatexEdit~\cite{ma2021latexify} and Image-to-LaTeX~\cite{gurgurov2024image}, focus on converting images of mathematical formulas into LaTeX code, generally involving lower task complexity. Generalist multimodal models, including Yi-VL~\cite{yivl}, Qwen2-VL~\cite{qwen2-vl}, Llama-3.2-11B-Vision-Instruct~\cite{llama3}, and Phi-3.5-Vision-Instruct~\cite{phi3}, leverage large-scale vision-language pre-training to generate code from image inputs across various domains. These models have demonstrated versatility in processing different types of visual data to produce code outputs. In a similar vein, closed-source multimodal models such as GPT-4o~\cite{gpt4}, Gemini-1.5-Pro~\cite{gemini}, and GLM-4-Plus~\cite{chatglm} exhibit strong performance in generating code from visual representations, including charts. However, existing works mainly focus on simple tasks that involve sophisticated logical relationships, such as flowcharts and architecture diagrams.

\section{Preliminaries}

We outline the foundational components of text-to-diagram system, detailing the processes involved in generating, coding, and editing diagrams based on user instructions.

\vspace{-4mm}
\paragraph{\ding{172} Diagram Generation}

Given an instruction \( x_{ins} \), our system generates code \( c_{diag} \), which can be compiled into a diagram \( D_{gen} \). Unlike standard code generation—which focuses on executable logic—our approach emphasizes both the logical structure and the aesthetic design.

Let \( f_{code}(x_{ins}) \) denote the function that translates a textual instruction \( x_{ins} \) into diagram-specific code \( c_{diag} \). Additionally, let \( f_{img}(c_{diag}) \) represent the function that compiles the diagram-specific code \( c_{diag} \) into the generated diagram \( D_{gen} \). The process can be formally expressed as follows:

\begin{equation}
D_{gen} = f_{img}(f_{code}(x_{ins})),
\label{eq:diagram_gen}
\end{equation}
where \( D_{gen} \) is the generated diagram based on the structured and logical elements specified in the user instruction.

\vspace{-4mm}
\paragraph{\ding{173} Diagram Coding} 

Given a diagram \( D_{gen} \), our system generates structured code \( c'_{diag} \) that accurately captures the logical framework and design elements of the diagram. This process ensures that each diagram not only conveys the intended information but also adheres to the necessary structural integrity. The process can be formulated as:
\begin{equation}
c'_{diag} = f_{code}^{-1}(D_{gen}).
\end{equation}

\vspace{-4mm}
\paragraph{\ding{174} Diagram Editing}

The process of modifying an existing diagram involves three main steps: \textit{diagram coding}, \textit{code modification}, and \textit{code-to-diagram} recompilation. Initially, the original diagram \( D_{ori} \) is translated back into its underlying code \( c_{ori} \). This code is subsequently updated according to the user's editing instruction \( x_{edit} \), resulting in the modified code \( c_{mod} \). Finally, this modified code \( c_{mod} \) is compiled into a new diagram \( D_{mod} \). The editing process can be mathematically represented as:

\begin{equation}
D_{mod} = f_{img}(f_{mod}(f_{code}^{-1}(D_{ori}), x_{edit})),
\label{eq:diagram_mod}
\end{equation}
where \( f_{code}^{-1}(D_{ori}) \) indicates the function that extracts the code \( c_{ori} \) from the original diagram \( D_{ori} \). The function \( f_{mod}(c_{ori}, x_{edit}) \) reflects the process of updating the code based on the instruction $x_{edit}$, ultimately producing the modified diagram \( D_{mod} \) after recompilation via \( f_{img} \).

\section{Method}

%下面这两块要不要显性说出来呢？还是在引言中说出来？
% \subsubsection{Distinction Between Diagrams and General Images}
% A crucial distinction in our work is the difference between \textit{natural images} and \textit{diagrams}. Natural images typically represent real-world scenes and do not necessarily require logical relationships. In contrast, diagrams are abstract visual representations that must clearly convey structured information and logical relationships between components.

% \subsubsection{Distinction Between Code and Diagram-specific Code}
% Similarly, there is a fundamental difference between \textit{code generation} and \textit{diagram-specific code generation}. Traditional code generation focuses on generating executable code that prioritizes functionality. However, diagram-specific code emphasizes logical accuracy and conceptual clarity, while also considering visual aesthetics such as node positioning, diagram layout, and labeling, which are critical in representing complex information effectively.

\begin{figure*}[htbp]
    \centering
    \includegraphics[width=\textwidth]{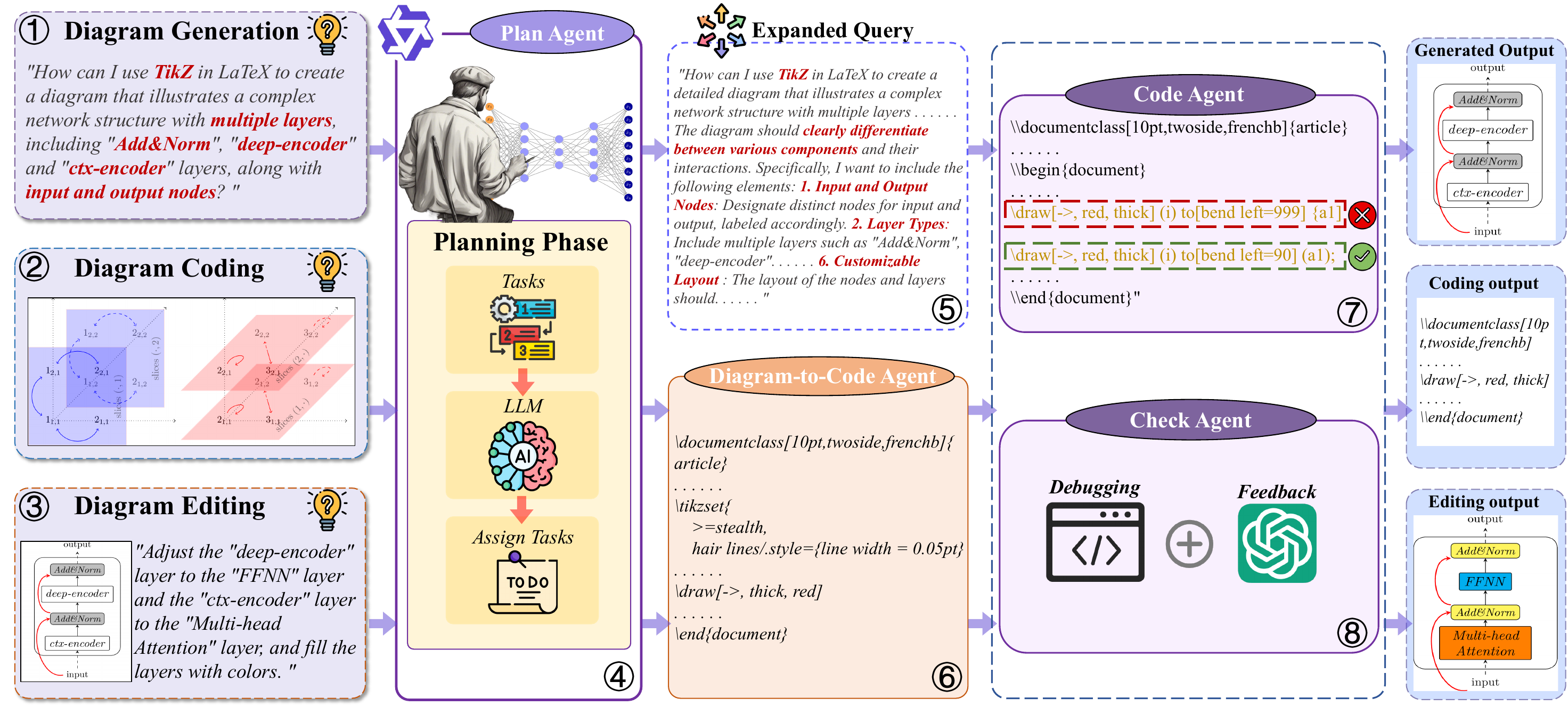}
    \caption{Workflow of DiagramAgent. The DiagramAgent handles diagram generation, coding, and editing tasks, processing the user query (\ding{172}-\ding{174}) through the Plan Agent (\ding{175}), which communicates with the Code Agent (\ding{178}) for diagram generation or with the Diagram-to-Code Agent (\ding{177}) for diagram coding or editing. Code is verified and feedback provided by the Check Agent (\ding{179}).}
    \vspace{-4mm}
    \label{fig:workflow}
\end{figure*}

We summarize DiagramAgent, which comprises four core agents: Plan Agent, Code Agent, Check Agent, and Diagram-to-Code Agent in Figure~\ref{fig:workflow}, highlighting the interactions between the agents and the processes involved in diagram generation, coding, and editing.

% \vspace{-1mm}
\subsection{Plan Agent}

The Plan Agent (\ding{175})~\cite{autoact,qwen25} is the central component responsible for interpreting user instructions and determining the necessary steps. The agent analyzes the instruction for completeness. If the instruction is incomplete (e.g., missing node or label information), the Plan Agent performs \textit{Query Expansion} using a Large Language Model (LLM), generating a more comprehensive query \( x_{comp} \), ensuring necessary elements are included. The process can be expressed as:
\begin{equation}
x_{comp} = f_{expand}(x_{ins})
\end{equation}
Next, the complete query \( x_{comp} \) is passed to the Code Agent, which generates the corresponding code \( c_{diag} \). 

For diagram coding, the Plan Agent handles the initial diagram \(D_{ori}\) by invoking the \textit{Diagram-to-Code Agent} \( f_{code}^{-1}(D_{ori}) \) (\ding{177}), which translates the diagram into its code form. For diagram editing, the Plan Agent receives the editing query \( x_{edit} \) along with the diagram \( D_{ori} \). Based on the user's request for changes, the Plan Agent again calls the \textit{Diagram-to-Code Agent} \( f_{code}^{-1}(D_{ori}) \) (\ding{177}) to convert the provided diagram into code. Then, the Plan Agent sends both the code and the editing query \( x_{edit} \) to the \textit{Code Agent} (\ding{178}).

% \vspace{-1mm}
\subsection{Code Agent}

The Code Agent (\ding{178})~\cite{unifying,qwen25} is responsible for transforming the processed user instructions from the Plan Agent (\ding{175}) into executable code. For diagram generation, the Code Agent takes the complete query \( x_{comp} \) and generates diagram-specific code \( c_{diag} \) as follows:
\begin{equation}
    c_{diag} = f_{code}(x_{comp})
\end{equation}
For diagram editing, the Code Agent takes the original code \( c_{ori} \) and applies the modifications according to the query \( x_{edit} \) to generate modified code \( c_{mod} \). The code is passed to the Check Agent (\ding{179}) for debugging and refinement, ensuring that the code is both syntactically and logically correct.

We optimize the Code Agent by training it to generate code \( c \) that closely matches a reference code \( c_{ref} \). This can be formalized as minimizing the following objective:
\begin{equation}
    \mathcal{L} = \min_{\theta} \mathbb{E}_{(x, c_{ref})} \left[ \mathcal{L}_{\text{code}}(f_{code_\theta}(x), c_{ref}) \right],
\end{equation}
where \( \theta \) represents the parameters of the model, and \( \mathcal{L}_{\text{code}} \) is the loss function, which measures the discrepancy between the generated code and the reference code.

\subsection{Diagram-to-Code Agent}
The Diagram-to-Code Agent (\ding{177}) is responsible for converting a given diagram \( D_{ori} \) into its corresponding code \( c'_{diag} \), facilitating the diagram coding task. The Diagram-to-Code Agent captures the diagram's logical structure and visual elements, translating them into structured code. This transformation process is formally expressed as:
\begin{equation}
    c'_{diag} = f_{\text{diagram-to-code}}(D_{ori})
\end{equation}
where \( c'_{diag} \) denotes the generated code based on the elements of the diagram $D_{ori}$.

The Diagram-to-Code Agent then sends the generated $c'_{diag}$ to the Check Agent (\ding{176}) for validation. If necessary, feedback is provided to refine \( c'_{diag} \). The objective is to generate code \( c'_{diag} \) that closely matches a reference code \( c_{ref} \), ensuring that the code preserves the diagram's logical and visual integrity. This optimization is formalized as minimizing the following loss function:
\begin{equation}
    \mathcal{L} = \min_{\theta} \mathbb{E}_{(D_{ori}, c_{ref})} \left[ \mathcal{L}_{\text{code}}(f_{\text{diagram-to-code}_\theta}(D_{ori}), c_{ref}) \right].
\end{equation}
where \( \mathcal{L}_{\text{code}} \) measures the discrepancy between the generated code and the reference code.
\subsection{Check Agent}
The Check Agent (\ding{179})~\cite{qwen25} is responsible for verifying the correctness and completeness of the code generated by the Code Agent (\ding{178}) and the Diagram-to-Code Agent(\ding{177}). Its role involves two critical stages: \textit{debugging} and \textit{verification}. First, the Check Agent compiles the generated code. If errors are detected, they are returned to the Code Agent or Diagram-to-Code Agent for correction and regeneration. Once the code compiles successfully, the Check Agent uses GPT-4o~\cite{gpt4} to verify the completeness of the code, ensuring that all necessary elements are included. The process can be summarized as follows:
% \begin{equation}
% f_{\text{check}}(c) = f_{\text{debug}}(c_{diag}, c_{mod}, c'_{diag}) + f_{\text{verify}}(c_{diag}, c_{mod}, c'_{diag})
% \end{equation}
{\small
\begin{equation}
f_{\text{check}}(c) = f_{\text{debug}}(c_{diag}, c_{mod}, c'_{diag}) + f_{\text{verify}}(c_{diag}, c_{mod}, c'_{diag})
\end{equation}
}

where \( f_{\text{debug}}(c) \) returns errors if present, and \( f_{\text{verify}}(c) \) evaluates the completeness of the diagram.

\vspace{-2mm}
\section{DiagramGenBenchmark}

We introduce \textit{DiagramGenBenchmark}, which focuses on transforming textual descriptions into structured diagram representations. It includes various types of diagrams such as model architecture diagrams, flowcharts, line charts, directed and undirected graphs, tables, bar charts, and mind maps. The data is sourced from HuggingFace's VGQA~\cite{vgbench}, datikz~\cite{AutomaTikZ}, and datikz-v2~\cite{AutomaTikZ} datasets, as well as open-source repositories on GitHub and Overleaf, which are licensed under CC BY 4.0 or MIT. These repositories predominantly feature diagram code written in LaTeX or DOT languages. Due to the limited space, we provide the benchmark dataset details in the Appendix~\ref{app:data_details}.

\begin{figure}[h]
    \centering
    \vspace{-3mm}
    \includegraphics[width=0.98\linewidth]{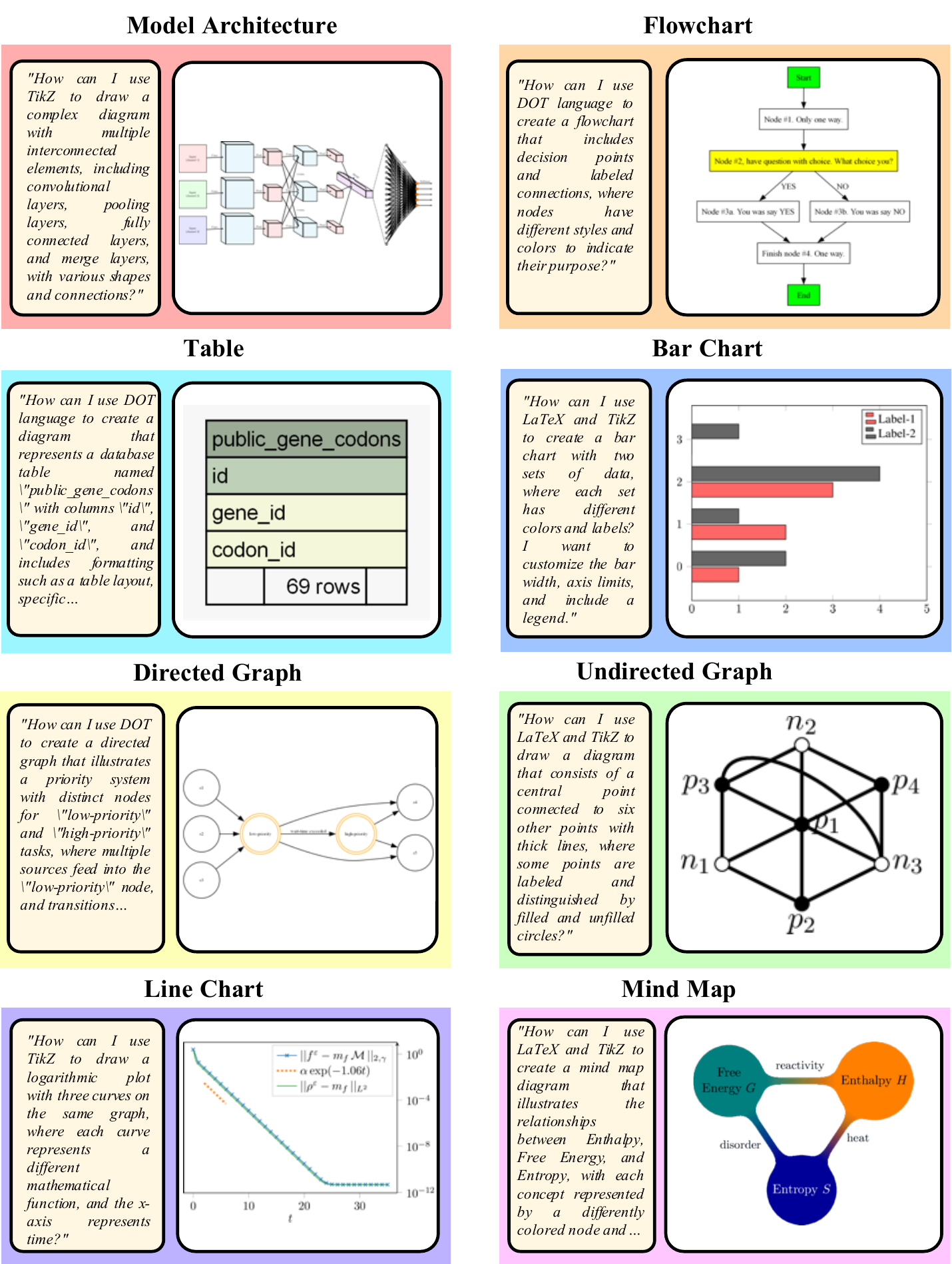}
    \caption{Example queries and diagrams.}
    \vspace{-3mm}
    \label{fig:data_examples}
\end{figure}

\vspace{-1mm}
\subsection{Data Statistics}
\vspace{-1mm}

Figure~\ref{fig:data_examples} provides example queries and diagrams, inlcuding model architecture, flowcharts, line charts, directed/undirected graphs, tables, bar charts, and mind maps. As detailed in Table~\ref{tab:data_statistics}, the diagram generation/coding task contains 6713 training samples and 270 testing samples, while the diagram editing task includes 1400 training samples and 200 testing samples. The query lengths and code lengths vary significantly, reflecting the complexity of the tasks.

\begin{table}[h]
    \vspace{-2mm}
    \footnotesize
    \centering
    \caption{Key statistics of DiagramGenBenchmark.}
    \vspace{-2mm}
    \label{tab:data_statistics}
    % \resizebox{\linewidth}{!}{
    \setlength{\tabcolsep}{0.6mm}{
    \begin{tabular}{lcc}
        \toprule
        \textbf{Statistic} & \textbf{Generation / Coding} & \textbf{Editing} \\
        \midrule
        Total Samples & 6983 & 1600 \\
        Training Samples & 6713 & 1400 \\
        Testing Samples & 270 & 200 \\
        \midrule
        \textbf{Diagram Type} & & \\
        \quad Model Architecture & 3000 (42.96\%) & 706 (44.13\%) \\
        \quad Flowchart & 1500 (21.48\%) & 320 (20\%) \\
        \quad Line Chart & 1142 (16.35\%) & 227 (14.19\%) \\
        \quad Directed Graph & 787 (11.27\%) & 217 (13.56\%) \\
        \quad Table & 114 (1.63\%) & 19 (1.19\%) \\
        \quad Bar Chart & 227 (3.25\%) & 77 (4.81\%) \\
        \quad Mind Map & 140 (2.00\%) & 16 (1\%) \\
        \quad Undirected Graph & 73 (1.05\%) & 18 (1.12\%) \\
        \midrule
        \textbf{Query Length (tokens)} & & \\
        \quad Minimum & 18 & 40 \\
        \quad Maximum & 561 & 499 \\
        \quad Average & 247 & 133 \\
        \midrule
        \textbf{Complete Query Length (tokens)} & & \\
        \quad Minimum & 404 & 115 \\
        \quad Maximum & 4015 & 3419 \\
        \quad Average & 1900 & 98 \\
        \midrule
        \textbf{Code Length (tokens)} & & \\
        \quad Minimum & 60 & 116 \\
        \quad Maximum & 75612 & 4727 \\
        \quad Average & 1701 & 1009.6 \\
        \bottomrule
    \end{tabular}}
    \vspace{-5mm}
\end{table}

% As shown in Table~\ref{tab:data_statistics}, the dataset encompasses a diverse range of diagram types, with the diagram generation / coding task containing 6713 training instances and 270 testing instances, and the diagram editing task comprising 1400 training instances and 200 testing instances. Each diagram type, such as model architecture diagrams, flowcharts, and directed graphs, is well-represented to ensure comprehensive evaluation across different diagram types.

% The query lengths vary significantly, with diagram generation / coding queries ranging from 18 to 561 tokens (avg. 247), while diagram editing task queries range from 40 to 499 tokens (avg. 133). Code lengths also display considerable variation, reflecting the complexity of diagrams, with diagram generation / coding length averaging at 1701 tokens and diagram editing task code length averaging at 1009.6 tokens.

\subsection{Evaluation Metrics}
% \vspace{-1mm}
\label{eval_metrics}
% DiagramGenBenchmark employs a range of evaluation metrics tailored to the specific requirements of diagram code generation and final output quality. 

\begin{itemize}
    \item \textbf{Code Generation Metrics}: For the diagram generation and editing tasks, we evaluate the generated code using:
        \begin{itemize}
            \item \textit{Pass@1} (↑): Measures accuracy of generated code.
            \item \textit{ROUGE-L} (↑): Measures similarity with reference code based on longest common subsequence.
            \item \textit{CodeBLEU} (↑): Assess semantic similarity with reference code.%\textit{BLEU} (↑)
            \item \textit{Edit Distance} (↓): Computes the edit operations required to match the reference.
            \item \textit{chrF} (↑),  \textit{RUBY} (↑): Evaluate text similarity and understanding. %\textit{bleurt} (↑),
        \end{itemize}
    \item \textbf{Diagram Quality Metrics}: For assessing the quality of generated diagrams, we use:
        \begin{itemize}
            % \item \textit{IS} (↑), \textit{FID} (↓), \textit{KID} (↓): Visual similarity metrics.
            \item \textit{CLIP-FID} (↓), \textit{LPIPS} (↓): Cross-modal similarity.
            \item \textit{PSNR} (↑), \textit{MS-SSIM} (↑): Structural and clarity metrics. %\textit{CMMD} (↓), \textit{SSIM} (↑), 
        \end{itemize}
    \item \textbf{Human Evaluation}: Three evaluators with graduate degrees assess the test set based on criteria rated from 1 to 5, with higher scores indicating better quality.
\end{itemize}

\begin{table*}[h]
    \centering
    \vspace{-2mm}
    \caption{Main results for diagram generation (Code Agent). The best result in each metric is bolded.}
    \vspace{-2mm}
    \label{tab:diagram generation-main}
    \resizebox{\textwidth}{!}{
    \begin{tabular}{l|c|c|c|c|c|c|c|c|c|c|c}
    \hline
    \textbf{Model} & \textbf{Size} & \textbf{Pass@1↑} & \textbf{ROUGE-L↑} & \textbf{CodeBLEU↑} & \textbf{Edit Dist.↓} & \textbf{chrF↑} & \textbf{RUBY↑} & \textbf{CLIP-FID↓} & \textbf{LPIPS↓} & \textbf{PSNR↑} & \textbf{MS-SSIM↑} \\
    \toprule
    Qwen2.5-Coder~\cite{qwen25} & 7B & 32.22 & 41.94 & 82.58 & 83.45 & 38.14 & 28.09 & 18.90 & 60.54 & 3.72 & 13.04 \\
    DeepSeek-Coder~\cite{deepseek} & 33B & 55.56 & 44.26 & 83.29 & 81.85 & 42.01 & 30.55 & 15.49 & 60.99 & 6.02 & 19.80 \\
    Code-Llama~\cite{codellama} & 34B & 8.89 & 22.92 & 76.78 & 95.89 & 28.77 & 13.60 & 30.12 & 59.80 & 0.89 & 2.32 \\
    WizardCoder~\cite{wizardcoder}, & 15B & 28.89 & 29.93 & 78.96 & 91.30 & 31.38 & 19.73 & 27.38 & 55.96 & 3.36 & 11.66 \\
    Codegeex4-all~\cite{zheng2023codegeex} & 9B & 49.63 & 42.14 & 82.94 & 86.31 & 41.36 & 28.69 & 13.86 & 61.08 & 5.48 & 17.37 \\
    Starcoder2~\cite{li2023starcoder} & 15B & 27.41 & 26.49 & 78.56 & 90.67 & 25.98 & 17.74 & 31.63 & 56.54 & 3.11 & 10.53 \\
    Yi-Coder~\cite{yicoder} & 9B & 37.04 & 41.38 & 82.46 & 83.91 & 39.20 & 28.00 & 22.40 & 57.10 & 3.91 & 14.11 \\
    \midrule
    Llama-3.1~\cite{llama3} & 8B & 33.58 & 37.04 & 80.45 & 88.24 & 36.80 & 24.79 & 17.91 & 58.80 & 3.78 & 11.94 \\
    Baichuan2~\cite{baichuan2} & 13B & 16.30 & 33.28 & 79.94 & 87.96 & 31.83 & 21.51 & 23.43 & 61.49 & 1.81 & 4.94 \\
    Internlm2\_5~\cite{internlm2} & 20B & 34.44 & 39.45 & 81.79 & 87.00 & 38.44 & 26.21 & 24.56 & 56.81 & 3.91 & 13.39 \\
    Yi-1.5~\cite{young2024yi} & 34B & 35.19 & 42.56 & 82.91 & 85.52 & 42.03 & 28.43 & 20.03 & 58.04 & 3.82 & 12.83 \\
    Qwen2~\cite{qwen25} & 7B & 41.48 & 41.74 & 82.49 & 84.86 & 39.72 & 27.93 & 15.57 & 58.89 & 4.60 & 15.48 \\
    \midrule
    GPT-4o~\cite{gpt4} & - & 49.81 & 44.59 & 82.83 & 85.17 & 43.83 & 30.08 & 13.26 & 63.07 & 5.56 & 18.21 \\
    DeepSeek V2.5~\cite{deepseekv2} & - & 54.44 & 43.00 & 82.83 & 85.67 & 43.63 & 28.75 & 13.32 & 62.32 & 5.56 & 16.98 \\
    GLM-4-plus~\cite{chatglm} & - & 42.96 & 46.42 & 83.91 & 82.40 & 44.51 & 32.13 & 14.70 & 63.38 & 4.47 & 13.89 \\
    Gemini~\cite{gemini} & - & 43.23 & 44.86 & 82.37 & 84.44 & 43.75 & 30.46 & 21.69 & 54.93 & 3.16 & 18.70 \\
    \midrule
    \textbf{DiagramAgent} & 7B & \textbf{58.15} & \textbf{51.97} & \textbf{86.83} & \textbf{74.62} & \textbf{53.49} & \textbf{39.71} & \textbf{11.16} & \textbf{45.95} & \textbf{6.38} & \textbf{24.78} \\
    \bottomrule
    \end{tabular}}
    \vspace{-1mm}
\end{table*}

\begin{table*}[h]
\centering
% \vspace{-1mm}
\caption{Ablation study for diagram generation task (Code Agent). Each result shows the performance of DiagramAgent Code Agent under various component configurations, with the decrease in each metric from the full model indicated in parentheses.}
\vspace{-2mm}
\label{tab:diagram generation-ablation}
\resizebox{\textwidth}{!}{
\begin{tabular}{l|c|c|c|c|c|c|c|c|c|c|c}
\toprule
\textbf{Diagram generation} & \textbf{Size} & \textbf{Pass@1↑} & \textbf{ROUGE-L↑} & \textbf{CodeBLEU↑} & \textbf{Edit Dist.↓} & \textbf{chrF↑} & \textbf{RUBY↑} & \textbf{CLIP-FID↓} & \textbf{LPIPS↓} & \textbf{PSNR↑} & \textbf{MS-SSIM↑} \\
\midrule
DiagramAgent & 7B & \textbf{58.15} & \textbf{51.97} & \textbf{86.83} & \textbf{74.62} & \textbf{53.49} & \textbf{39.71} & \textbf{11.16} & \textbf{45.95} & \textbf{6.38} & \textbf{24.78} \\
\midrule
-- w/o GPT-4o & 7B & 57.78 & 51.95 & 86.35 & 74.71 & 52.81 & 39.33 & 11.23 & 46.66 & 6.33 & 20.80 \\
& & (-0.37) & (-0.02) & (-0.48) & (+0.09) & (-0.68) & (-0.38) & (+0.07) & (+0.71) & (-0.05) & (-3.98) \\
-- w/o Compiler & 7B & 57.41 & 51.94 & 85.78 & 74.68 & 51.74 & 39.65 & 11.29 & 48.13 & 5.93 & 24.10 \\
& & (-0.74) & (-0.03) & (-1.05) & (+0.06) & (-1.75) & (-0.06) & (+0.13) & (+2.18) & (-0.45) & (-0.68) \\
-- w/o GPT-4o \& Compiler & 7B & 57.41 & 51.87 & 85.74 & 74.79 & 51.69 & 38.77 & 11.30 & 48.17 & 5.85 & 20.37 \\
& & (-0.74) & (-0.10) & (-1.09) & (+0.17) & (-1.80) & (-0.94) & (+0.14) & (+2.22) & (-0.53) & (-4.41) \\
\bottomrule
\end{tabular}}
\vspace{-2mm}
\end{table*}

\section{Experiment}
% \subsection{Dataset}
% 数据集包括 
% diagram generation、
% image-to-code、
% 修改指令 + code-to-new_code 三部分数据集
% 数据集构建图
% 数据分布表
% 数据集示例图 把生图放到正文中，把改图放到附录中，两张图
\vspace{-1mm}
\paragraph{Setup}
DiagramAgent performs diagram generation and editing through four core agents: Plan Agent, Code Agent, Check Agent, and Diagram-to-Code Agent. The Plan Agent utilizes the Qwen-72B~\cite{qwen25} to effectively distinguish and interpret the user's query tasks. The Code Agent is based on Qwen2.5-Coder-7B~\cite{qwen25}, fine-tuned for four epochs with a maximum input length of 8192 tokens. Diagram-to-Code Agent uses Qwen2-VL-7B~\cite{qwen2-vl} in its default configuration. Experiments are conducted on an 8$\times$80G A100 GPUs.

\vspace{-4mm}
\paragraph{Model Comparison}
We evaluate DiagramAgent's Code Agent against 16 mainstream models. The specialized code models include Qwen2.5-Coder~\cite{qwen25}, DeepSeek-Coder-Instruct~\cite{deepseek}, Code-Llama~\cite{codellama}, WizardCoder~\cite{wizardcoder}, Codegeex4-all~\cite{zheng2023codegeex}, Starcoder2~\cite{li2023starcoder}, and Yi-Coder~\cite{yicoder}. Open-source models feature Llama-3.1~\cite{llama3}, Baichuan2~\cite{baichuan2}, Internlm2\_5~\cite{internlm2}, Yi-1.5~\cite{young2024yi}, and Qwen2~\cite{qwen25}, while close-source models include GPT-4o~\cite{gpt4}, DeepSeek V2.5~\cite{deepseekv2}, GLM-4-plus~\cite{chatglm}, and Gemini~\cite{gemini}. Similarly, we compare the Diagram-to-Code Agent's performance on the diagram coding against open-source models include Yi-VL~\cite{yivl}, Qwen2-VL~\cite{qwen2-vl}, Internlm-Xcomposer2.5~\cite{zhang2024internlm}, Llama-3.2-Vision~\cite{llama3}, Phi-3.5-vision~\cite{phi3}, Llava-v1.6~\cite{liu2024llavanext}, Cogvlm2-llama3~\cite{hong2024cogvlm2}, and Deepseek-vl~\cite{deepseekvl}. Close-source models include GPT-4o~\cite{gpt4}, GLM-4-plus~\cite{chatglm}, and Gemini-1.5-pro~\cite{gemini}. 

\subsection{Diagram Generation}

\paragraph{Main Results}
DiagramAgent's Code Agent demonstrates outstanding performance on the diagram generation task, achieving top scores across both code accuracy and image fidelity metrics, as shown in Table \ref{tab:diagram generation-main}. In terms of code quality, DiagramAgent achieves leading results with metrics such as Pass@1 (58.15), ROUGE-L (51.97), and CodeBLEU (86.83), among others, highlighting its capability to generate accurate and robust code representations. These results demonstrate the effectiveness of DiagramAgent in generating structured, accurate, and high-quality diagrams. For image quality, it also excels with PSNR (6.38), LPIPS (45.95), and others, confirming its ability to maintain high visual fidelity in generated diagrams.

\vspace{-4mm}
\paragraph{Ablation Study}
The ablation study in Table \ref{tab:diagram generation-ablation} demonstrates the critical role of the Check Agent in enhancing DiagramAgent's performance. Removing the verification module (GPT-4o) leads to the chrF decrease of -0.68 and the LPIPS increase of +0.71. Excluding the debugging component (compiler) results in a significant chrF drop of -1.75 and an LPIPS increase of +2.18, impacting both code quality and image fidelity. Together, the debugging and verification achieve the highest scores across metrics, underscoring their combined value in producing precise diagrams. Overall, this ablation study demonstrates that both debugging and verification are essential in diagram generation.

\subsection{Diagram Coding}
\label{Diagram_Coding}

\paragraph{Main Results} DiagramAgent's model, configured with compiler-based debugging followed by GPT-4o verification, achieves the highest performance across several metrics in the diagram coding task, as illustrated in Table \ref{tab:image-to-code-main}. Key metrics include Pass@1 (68.89), ROUGE-L (48.99), codeBLEU (84.64), demonstrating DiagramAgent's effectiveness in generating high-quality code from images. Compared to both open-source models like Qwen2-VL-7B-Instruct and closed-source models such as gpt-4o, DiagramAgent consistently excels, highlighting its robustness in tasks requiring precise visual-to-code translation. 

\begin{table}[!h]
\centering
\begin{minipage}[!t]{\columnwidth}
    \vspace{-2mm}
    \footnotesize
    % \centering
    \caption{Main results for diagram coding task (Diagram-to-Code Agent). The best result in each metric is bolded.}
    \vspace{-2mm}
    \label{tab:image-to-code-main}
    % \resizebox{\linewidth}{!}{
    \setlength{\tabcolsep}{0.7mm}{
    \begin{tabular}{l|c|c|c|c}
    \toprule
    \textbf{Model} & \textbf{Size} & \textbf{Pass@1↑} & \textbf{ROUGE-L↑} & \textbf{codeBLEU↑} \\
    \midrule
    Yi-VL~\cite{yivl} & 34B & 2.22 & 20.01 & 70.57 \\
    Qwen2-VL~\cite{qwen2-vl} & 7B & 28.89 & 31.74 & 80.04 \\
    Internlm-xcomposer2.5~\cite{zhang2024internlm} & 7B & 3.33 & 28.47 & 77.35 \\
    Llama-3.2-Vision~\cite{llama3} & 11B & 27.78 & 21.94 & 75.37 \\
    Phi-3.5-vision~\cite{phi3} & 4B & 24.07 & 27.53 & 76.56 \\
    Llava-v1.6~\cite{liu2024llavanext} & 34B & 8.89 & 26.68 & 76.53 \\
    Cogvlm2-llama3~\cite{hong2024cogvlm2} & 19B & 3.70 & 14.42 & 70.72 \\
    Deepseek-vl~\cite{deepseekvl} & 7B & 50.74 & 25.18 & 76.48 \\
    \midrule
    GPT-4o~\cite{gpt4} & - & 64.07 & 39.95 & 81.78 \\
    GLM-4-plus~\cite{chatglm} & - & 51.48 & 35.92 & 80.16 \\
    Gemini-1.5-pro~\cite{gemini} & - & 17.78 & 38.66 & 80.75 \\
    \midrule
    \textbf{DiagramAgent} & 7B & \textbf{68.89} & \textbf{48.99} & \textbf{84.64} \\
    \bottomrule
    \end{tabular}}
    \vspace{0.5mm}
\end{minipage}

% \begin{table}[!h]
\begin{minipage}[!b]{\columnwidth}
    \centering
    \vspace{-1mm}
    \caption{Ablation study for diagram coding task (Diagram-to-Code Agent). Each result shows the performance of DiagramAgent under various component configurations, with the decrease in each metric from the full model indicated in parentheses.}
    \vspace{-2mm}
    \label{tab:image-to-code-ablation}
    \resizebox{\linewidth}{!}{
    \begin{tabular}{l|c|c|c|c}
    \toprule
    \textbf{Diagram coding} & \textbf{Size} & \textbf{Pass@1↑} & \textbf{ROUGE-L↑} & \textbf{codeBLEU↑}\\
    \midrule
    DiagramAgent & 7B & \textbf{68.89} & \textbf{48.99} & \textbf{84.64} \\
    \midrule
    -- w/o GPT-4o & 7B & 62.59 & 48.71 & 84.57\\
    & & (-6.30) & (-0.28) & (-0.07) \\
    -- w/o Compiler & 7B & 53.33 & 48.46 & 84.52  \\
    & & (-15.56) & (-0.53) & (-0.12) \\
    -- w/o GPT-4o \& Compiler & 7B & 52.59 & 47.81 & 84.21  \\
    & & (-16.30) & (-1.18) & (-0.43) \\
    \bottomrule
    \end{tabular}}
    % \vspace{-6mm}
\end{minipage}
\vspace{-7mm}
\end{table}
% \vspace{-6mm}
% \end{table}

% \begin{table}[!h]
% \centering
% \vspace{-2mm}
% \caption{Ablation study for diagram coding task (Diagram-to-Code Agent). Each result shows the performance of DiagramAgent under various component configurations, with the decrease in each metric from the full model indicated in parentheses.}
% \vspace{-2mm}
% \label{tab:image-to-code-ablation}
% \resizebox{\linewidth}{!}{
% \begin{tabular}{l|c|c|c|c}
% \toprule
% \textbf{Diagram coding} & \textbf{Size} & \textbf{Pass@1↑} & \textbf{ROUGE-L↑} & \textbf{codeBLEU↑}\\
% \midrule
% DiagramAgent & 7B & \textbf{68.89} & \textbf{48.99} & \textbf{84.64} \\
% \midrule
% -- w/o GPT-4o & 7B & 62.59 & 48.71 & 84.57\\
% & & (-6.30) & (-0.28) & (-0.07) \\
% -- w/o Compiler & 7B & 53.33 & 48.46 & 84.52  \\
% & & (-15.56) & (-0.53) & (-0.12) \\
% -- w/o GPT-4o \& Compiler & 7B & 52.59 & 47.81 & 84.21  \\
% & & (-16.30) & (-1.18) & (-0.43) \\
% \bottomrule
% \end{tabular}}
% \vspace{-6mm}
% \end{table}

\paragraph{Ablation Study} 

The ablation study in Table \ref{tab:image-to-code-ablation} demonstrates the effectiveness and necessity of the \textbf{Check Agent}’s debugging (compiler) and verification (GPT-4o) components within DiagramAgent. Removing the verification module results in a notable Pass@1 reduction of -6.30 and a ROUGE-L reduction of -0.28, underscoring the importance of verification for accurate code generation. Excluding the compiler module reduces Pass@1 by -15.56 and ROUGE-L by -0.53. 
The full model configuration, which includes both debugging and verification, achieves optimal results across all metrics, affirming the essential role of these components in enhancing diagram coding performance.

Due to space limitations, the full results presented in Table~\ref{tab:image-to-code-main} and Table~\ref{tab:image-to-code-ablation} are provided in Appendix~\ref{app:image-to-code-complete}.

% \begin{table}[!h]
% \centering
% \vspace{-2mm}
% \caption{Ablation study for diagram coding task (Diagram-to-Code Agent). Each result shows the performance of DiagramAgent under various component configurations, with the decrease in each metric from the full model indicated in parentheses.}
% \vspace{-2mm}
% \label{tab:image-to-code-ablation}
% \resizebox{\linewidth}{!}{
% \begin{tabular}{l|c|c|c|c}
% \toprule
% \textbf{Diagram coding} & \textbf{Size} & \textbf{Pass@1↑} & \textbf{ROUGE-L↑} & \textbf{codeBLEU↑}\\
% \midrule
% DiagramAgent & 7B & \textbf{68.89} & \textbf{48.99} & \textbf{84.64} \\
% \midrule
% -- w/o GPT-4o & 7B & 62.59 & 48.71 & 84.57\\
% & & (-6.30) & (-0.28) & (-0.07) \\
% -- w/o Compiler & 7B & 53.33 & 48.46 & 84.52  \\
% & & (-15.56) & (-0.53) & (-0.12) \\
% -- w/o GPT-4o \& Compiler & 7B & 52.59 & 47.81 & 84.21  \\
% & & (-16.30) & (-1.18) & (-0.43) \\
% \bottomrule
% \end{tabular}}
% \vspace{-6mm}
% \end{table}

\begin{table*}[!h]
\centering
\caption{Main results for diagram editing (Code Agent). The best result in each metric is bolded.}
\label{tab:modify-diagram generation-main}
\resizebox{\textwidth}{!}{
\begin{tabular}{l|c|c|c|c|c|c|c|c|c|c|c}
\toprule
\textbf{Model} & \textbf{Size} & \textbf{Pass@1↑} & \textbf{ROUGE-L↑} & \textbf{CodeBLEU↑} & \textbf{Edit Dist.↓} & \textbf{chrF↑} & \textbf{RUBY↑} & \textbf{CLIP-FID↓} & \textbf{LPIPS↓} & \textbf{PSNR↑} & \textbf{MS-SSIM↑} \\
\midrule
Qwen2.5-Coder-7B~\cite{qwen25} & 7B & 71.50 & 91.86 & 97.42 & 13.26 & 89.91 & 86.99 & 4.79 & 46.45 & 11.16 & 66.76 \\
DeepSeek-Coder-Instruct~\cite{deepseek} & 33B & 90.50 & 96.64 & 98.48 & 5.80 & 95.73 & 94.68 & 2.63 & 46.25 & 15.84 & 86.42 \\
Code-Llama~\cite{codellama} & 34B & 87.00 & 52.51 & 92.55 & 50.96 & 65.83 & 40.22 & 4.95 & 44.62 & 24.10 & 82.42 \\
WizardCoder~\cite{wizardcoder} & 15B & 87.50 & 74.59 & 95.20 & 28.91 & 84.23 & 63.71 & 4.92 & 44.18 & \textbf{24.24} & 82.88 \\
Codegeex4-all~\cite{zheng2023codegeex} & 9B & 90.00 & 96.73 & 98.71 & 5.39 & 95.99 & 95.69 & 1.93 & 43.43 & 11.47 & 92.35 \\
Starcoder2~\cite{li2023starcoder} & 15B & 41.00 & 21.34 & 90.28 & 80.79 & 34.04 & 14.36 & 9.44 & 44.76 & 11.50 & 37.93 \\
Yi-Coder~\cite{yicoder} & 9B & 81.50 & 96.03 & 98.08 & 7.00 & 95.43 & 93.41 & 2.68 & 45.28 & 13.05 & 78.59 \\
\midrule
Llama-3.1-8B-Instruct~\cite{llama3} & 8B & 24.00 & 50.85 & 89.20 & 55.86 & 57.03 & 44.52 & 14.06 & 48.59 & 4.09 & 21.37 \\
Baichuan2-13B-Chat~\cite{baichuan2} & 13B & 39.50 & 82.07 & 92.51 & 30.60 & 82.33 & 75.16 & 10.04 & 44.80 & 6.50 & 37.06 \\
Internlm2\_5-20b-chat~\cite{internlm2} & 20B & 57.00 & 84.31 & 95.57 & 21.13 & 87.98 & 77.90 & 6.58 & 43.85 & 12.14 & 55.96 \\
Yi-1.5-34B-chat~\cite{young2024yi} & 34B & 90.50 & 96.64 & 98.38 & 6.85 & 95.78 & 94.52 & 2.13 & 45.86 & 16.06 & 85.70 \\
Qwen2-7B-Instruct~\cite{qwen25} & 7B & 81.50 & 91.51 & 96.40 & 17.87 & 91.34 & 87.63 & 3.72 & 44.87 & 15.59 & 76.67 \\
\midrule
GPT-4o~\cite{gpt4} & - & 92.42 & 96.22 & 97.73 & 7.31 & 95.49 & 94.50 & 1.89 & 43.53 & 14.23 & 88.43 \\
DeepSeek V2.5~\cite{deepseekv2} & - & 95.00 & 96.77 & 98.83 & 5.04 & 96.10 & 94.96 & 1.63 & 43.16 & 12.81 & 91.97 \\
GLM-4-plus~\cite{chatglm} & - & 92.00 & 97.05 & 98.63 & 6.06 & 96.04 & 95.12 & 1.54 & 45.79 & 13.89 & 88.31 \\
Gemini~\cite{gemini} & - & 72.00 & 95.09 & 95.34 & 7.00 & 93.32 & 93.45 & 2.08 & 47.57 & 12.50 & 85.59 \\
\midrule
\textbf{DiagramAgent Code Agent} & 7B & \textbf{98.00} & \textbf{98.41} & \textbf{99.93} & \textbf{3.58} & \textbf{97.96} & \textbf{97.05} & \textbf{1.08} & \textbf{40.64} & 13.18 & \textbf{97.00} \\
\bottomrule
\end{tabular}}
% \vspace{-2mm}
\end{table*}

\begin{table*}[h]
    \centering
    \caption{Ablation study for diagram editing task (Code Agent). Each result shows the performance of DiagramAgent Code Agent under various component configurations, with the decrease in each metric from the full model indicated in parentheses.}
    \label{tab:modify-diagram generation-ablation}
    \resizebox{\textwidth}{!}{
    \begin{tabular}{l|c|c|c|c|c|c|c|c|c|c|c}
    \toprule
    \textbf{Diagram editing} & \textbf{Size} & \textbf{Pass@1↑} & \textbf{ROUGE-L↑} & \textbf{CodeBLEU↑} & \textbf{Edit Dist.↓} & \textbf{chrF↑} & \textbf{RUBY↑} & \textbf{CLIP-FID↓} & \textbf{LPIPS↓} & \textbf{PSNR↑} & \textbf{MS-SSIM↑} \\
    \midrule
    DiagramAgent & 7B & \textbf{98.00} & \textbf{98.41} & \textbf{99.93} & \textbf{3.58} & \textbf{97.96} & \textbf{97.05} & \textbf{1.08} & \textbf{40.64} & 13.18 & \textbf{97.00} \\
    \midrule
    -- w/o GPT-4o & 7B & 97.00 & 98.19 & 99.29 & 3.63 & 97.88 & 96.96 & 1.13 & 41.95 & 13.33 & 96.02 \\
    & & (-1.00) & (-0.22) & (-0.64) & (+0.05) & (-0.08) & (-0.09) & (+0.05) & (+1.31) & (+0.15) & (-0.98) \\
    -- w/o Compiler & 7B & 95.50 & 98.12 & 99.19 & 3.84 & 97.45 & 96.91 & 1.26 & 42.17 & \textbf{16.72} & 96.22 \\
    & & (-2.50) & (-0.29) & (-0.74) & (+0.26) & (-0.51) & (-0.14) & (+0.18) & (+1.53) & (+3.54) & (-0.78) \\
    -- w/o GPT-4o \& Compiler & 7B & 95.50 & 97.92 & 99.11 & 3.96 & 97.30 & 96.73 & 1.36 & 42.21 & 15.60 & 93.93 \\
    & & (-2.50) & (-0.49) & (-0.82) & (+0.38) & (-0.66) & (-0.32) & (+0.28) & (+1.57) & (+2.42) & (-3.07) \\
    \bottomrule
    \end{tabular}}
\end{table*}

\subsection{Diagram Editing}

\paragraph{Main results} 
DiagramAgent’s Code Agent achieves superior results on the diagram editing task, demonstrating exceptional performance in both code accuracy and image quality metrics, as shown in Table \ref{tab:modify-diagram generation-main}. In terms of code generation, DiagramAgent leads with top scores in Pass@1 (98.00), ROUGE-L (98.41), and CodeBLEU (99.93), underscoring its capability for precise and reliable code outputs. For image quality, DiagramAgent also excels, achieving outstanding results in CLIP-FID (1.08), LPIPS (40.64), and MS-SSIM (97.00), although its PSNR (13.18) is slightly lower than WizardCoder, potentially due to DiagramAgent's focus on overall image fidelity rather than absolute sharpness. DiagramAgent consistently outperforms baseline models across comprehensive metrics, validating its strong adaptability and reliability.

\paragraph{Ablation Study} 

As shown in Table~\ref{tab:modify-diagram generation-ablation}, removing the verification module results in a Pass@1 decrease of -1.00 and a chrF reduction of -0.08, highlighting its role in ensuring code accuracy. Excluding the debugging module notably impacts CodeBLEU by -0.74 and increases LPIPS by +1.53, affecting both code quality and visual fidelity. Interestingly, while removing the debugging module led to a decrease in overall fidelity, PSNR showed a slight improvement, possibly due to the model's focus shifting towards simpler representations, which, though less precise, may result in visually sharper images. Removing both components leads to larger performance declines, including a Pass@1 reduction of -2.50 and an MS-SSIM decrease of -3.07, confirming the necessity of debugging and verification for maintaining both high accuracy and fidelity.

\subsection{Human Evaluation}

\begin{figure}[h]
    \centering
    \includegraphics[width=0.8\linewidth]{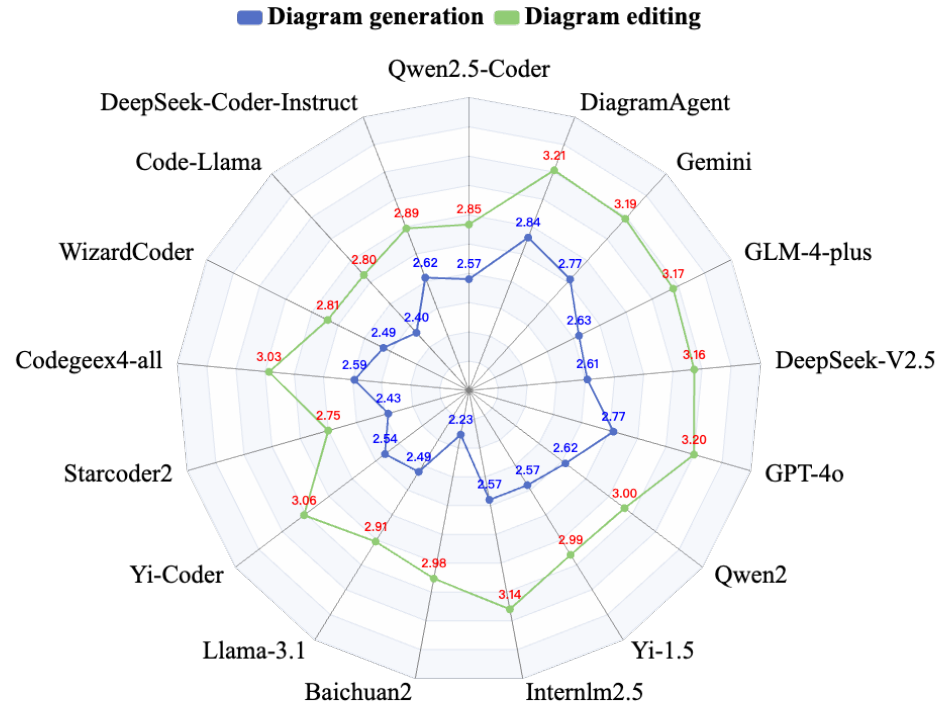} 
    \caption{Human evaluation results for different models on diagram generation and Modify diagram generation tasks.}
    \label{fig:human-eval}
    \vspace{-4mm}
\end{figure}

To assess DiagramAgent’s performance, three raters with master’s degrees evaluated two core tasks—the diagram generation task and the diagram editing task—using a 1-5 scale. A score of 1 represents minimal satisfaction in terms of code accuracy and diagram fidelity, while a score of 5 indicates the highest satisfaction. The evaluation criteria focused on the similarity between the generated diagram and the reference diagram.
Figure~\ref{fig:human-eval} presents the scores across models. DiagramAgent achieves state-of-the-art scores in both tasks, demonstrating superior accuracy compared to other models. Notably, the results show that DiagramAgent performs better on the diagram editing task than on the diagram generation task. This difference suggests that diagram editing task, which involves modifying an existing codebase, is comparatively easier due to the established structure, whereas the diagram generation task requires generating a diagram from scratch, posing a higher challenge given the dependency on instruction quality. 

Consequently, achieving accurate diagrams in a single generation step remains a significant challenge, often requiring multiple iterations to refine the output. This observation highlights the importance of DiagramAgent's capability for iterative diagram editing, enabling the generation of precise and refined diagrams. We also provide detailed error analysis in Appendix~\ref{app:error-analysis}.

% \subsection{Error Analysis}
% Our error analysis covers three tasks: diagram generation, diagram coding, and diagram editing, as shown in Figure \ref{fig:error-analysis}.
% For the diagram generation task, errors typically involve omissions in transition labels and node placements. The example reveals minor label and positioning discrepancies, which, although subtle, reduce diagram accuracy and clarity. For the Modify diagram generation task, color adjustments are prone to errors. As seen, while the instruction specified blue for certain lines, the model incorrectly applied red. Such color mismatches impact visual coherence, especially when colors are semantically meaningful. For the diagram coding task, inaccuracies often arise in attributes like font size and font style. In the example, misinterpretations of font and color settings lead to visual deviations from the ground truth, impacting fidelity. Further detailed error patterns are discussed in Appendix X.

\section{Conclusion and Limitations}

In this work, we present DiagramAgent, a framework for Text-to-Diagram generation, alongside DiagramGenBenchmark, a comprehensive dataset that encompasses eight distinct types of structured diagrams. Our extensive experimental evaluations, which include both objective metrics and human assessments, demonstrate that DiagramAgent significantly outperforms existing open-source and close-source models in diagram generation, coding, and editing tasks. To the best of our knowledge, this is the first framework specifically designed not only for generating structured diagrams from textual descriptions but also for editing them. This work sets a new benchmark for text-to-diagram tasks, contributing significantly to the fields of structured diagram generation and editing.

% 上面这句表述不知道准不准确，还是更进一步细化表述呢？
However, there are still areas for further refinement. DiagramAgent's handling of more complex structures, particularly with regard to intricate spatial relationships, remains an area for future development. Additionally, while diagram editing has reached a high level of accuracy, challenges persist in diagram generation, where achieving consistent and precise coding remains critical for producing reliable and accurate modifications.
% Despite these advancements, limitations remain. DiagramAgent's ability to handle complex structures, particularly intricate spatial relationships, requires further improvement. Additionally, effective diagram editing faces challenges due to the need for higher diagram coding accuracy. This work establishes a new benchmark for Text-to-Diagram tasks and contributes to the structured diagram generation and editing.

{
    \small
    \bibliographystyle{ieeenat_fullname}
    \bibliography{ref}
}

\input{appendix}

\end{document}

%% file: appendix.tex
\clearpage
\setcounter{page}{1}
\maketitlesupplementary
\appendix
% \immediate\write\@auxout{\string\let\string\contentsline\string\gobble} 
% \immediate\write\@auxout{\string\def\string\addcontentsline\string#1\string#2\string#3{}}

% \tableofcontents

\begin{figure*}[!ht]
    \centering
    \includegraphics[width=\linewidth]{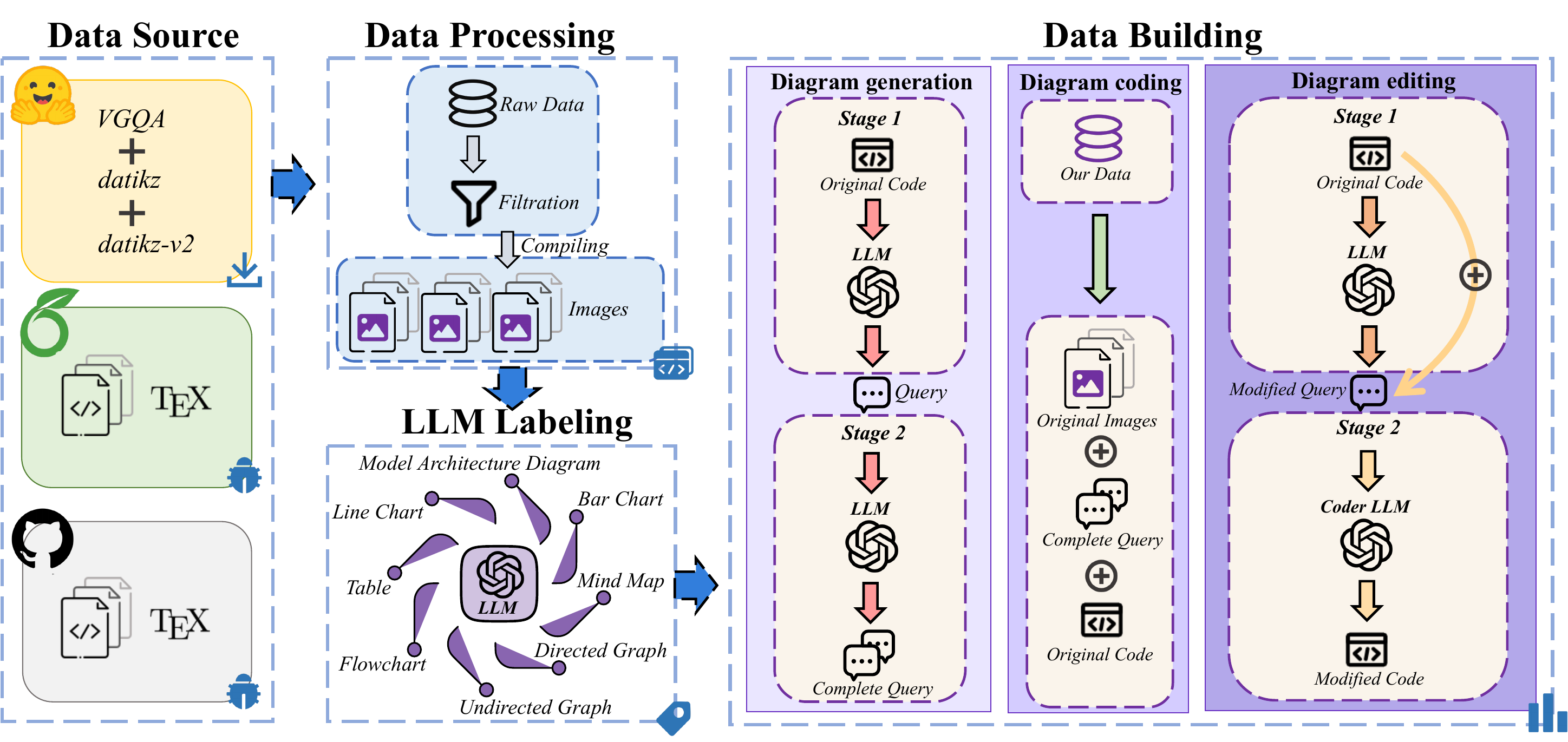}
    \caption{DiagramGenBenchmark Data Curation Process}
    \label{fig:data_curation}
\end{figure*}

\section{DiagramGenBenchmark Details}
\label{app:data_details}

\subsection{Dataset Curation}
\label{app:data_curation}

In constructing DiagramGenBenchmark, we employed a rigorous data curation process to ensure quality and diversity across the dataset. This section details our multi-step pipeline, from raw data collection to task-specific dataset creation, as illustrated in Figure~\ref{fig:data_curation}.

\paragraph{Data Sources} We gathered raw Tex and DOT language-based code data from multiple sources, including the VGQA~\cite{vgbench}, datikz~\cite{AutomaTikZ}, and datikz-v2~\cite{AutomaTikZ} datasets on Hugging Face, as well as publicly available repositories on GitHub and Overleaf. These sources yielded a comprehensive collection of over 13,000 raw code samples, originally sourced from various research papers on arXiv.

\paragraph{Data Processing} Collected raw code samples were subjected to manual filtering. Out of 13,000 initial samples, we retained only those that could be compiled successfully into images, ensuring code-to-diagram consistency. We executed each code sample using respective Tex and DOT compilers and stored each code-image pair in separate directories. After this filtering, we obtained 6,983 code-image pairs as the final dataset.

\paragraph{Data Annotation} To label the data, we defined eight diagram categories: model architecture diagram, flowchart, line chart, directed graph, undirected graph, table, bar chart, and mind map. Automatic labeling was conducted using the GPT-4o~\cite{gpt4}.

\subsection{Dataset Task}

\textbf{Dataset Tasks}: We constructed three distinct tasks to enhance the benchmark's applicability:
\begin{itemize}
    \item \textit{Diagram generation}: This task involves generating code from human instructions. We created human-like instructions by reverse-engineering existing open-source diagram code using the closed-source GPT-4o model. A total of 6,713 training instances and 270 testing instances were generated for this task.
    \item \textit{Diagram coding}: This task requires generating compliable diagram code from an image. Using the code and image pairs from the diagram generation task, we retained only image-code pairs and created a set of 6,713 training instances and 270 testing instances.
    \item \textit{Diagram editing}: This task simulates code editing based on revision instructions. We generated modification suggestions using GPT-4o and applied them to the original code with the CodeGeeX-4~\cite{zheng2023codegeex} model. Only compilable, unique modified codes were selected, resulting in 1,400 training instances and 200 testing instances.
\end{itemize}

\subsection{Dataset Distribution}

\begin{figure*}[h]
    \centering
    \includegraphics[width=\linewidth]{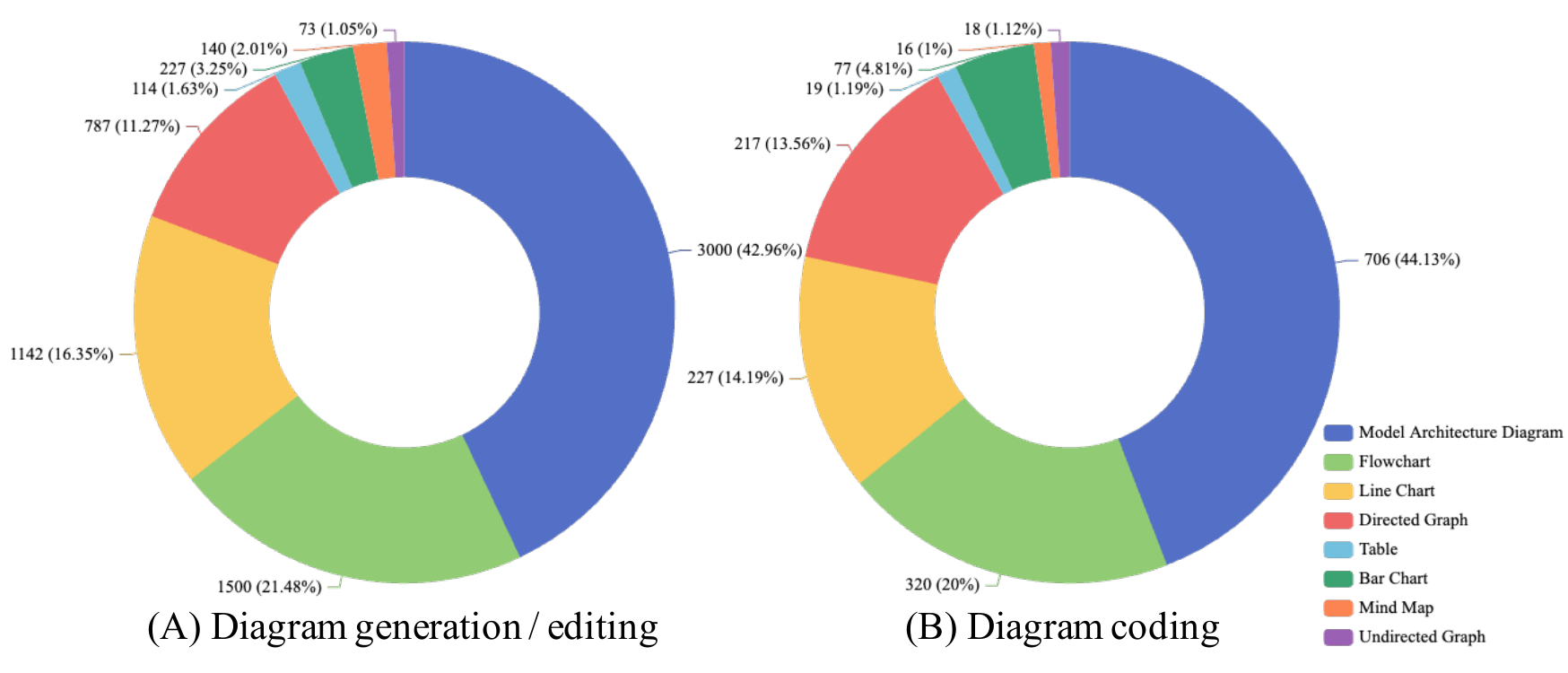}
    \caption{Data Distribution of DiagramGenBenchmark.}
    \label{fig:data_distribution}
\end{figure*}

Figure~\ref{fig:data_distribution} illustrates the data distribution for different diagram types in the diagram generation / diagram coding task and the diagram editing task.

\section{Detailed Prompts for DiagramAgent}
This appendix provides a comprehensive view of the prompt designs used in DiagramAgent's core components, including the Plan Agent, Complete Query handling, Diagram-to-Code Agent, Code Agent, and Check Agent. These prompts support different tasks such as drawing, modifying diagrams, and error checking, ensuring that each module contributes to accurate and coherent diagram generation and modification.

\begin{figure}[ht]
    \centering
    \includegraphics[width=\linewidth]{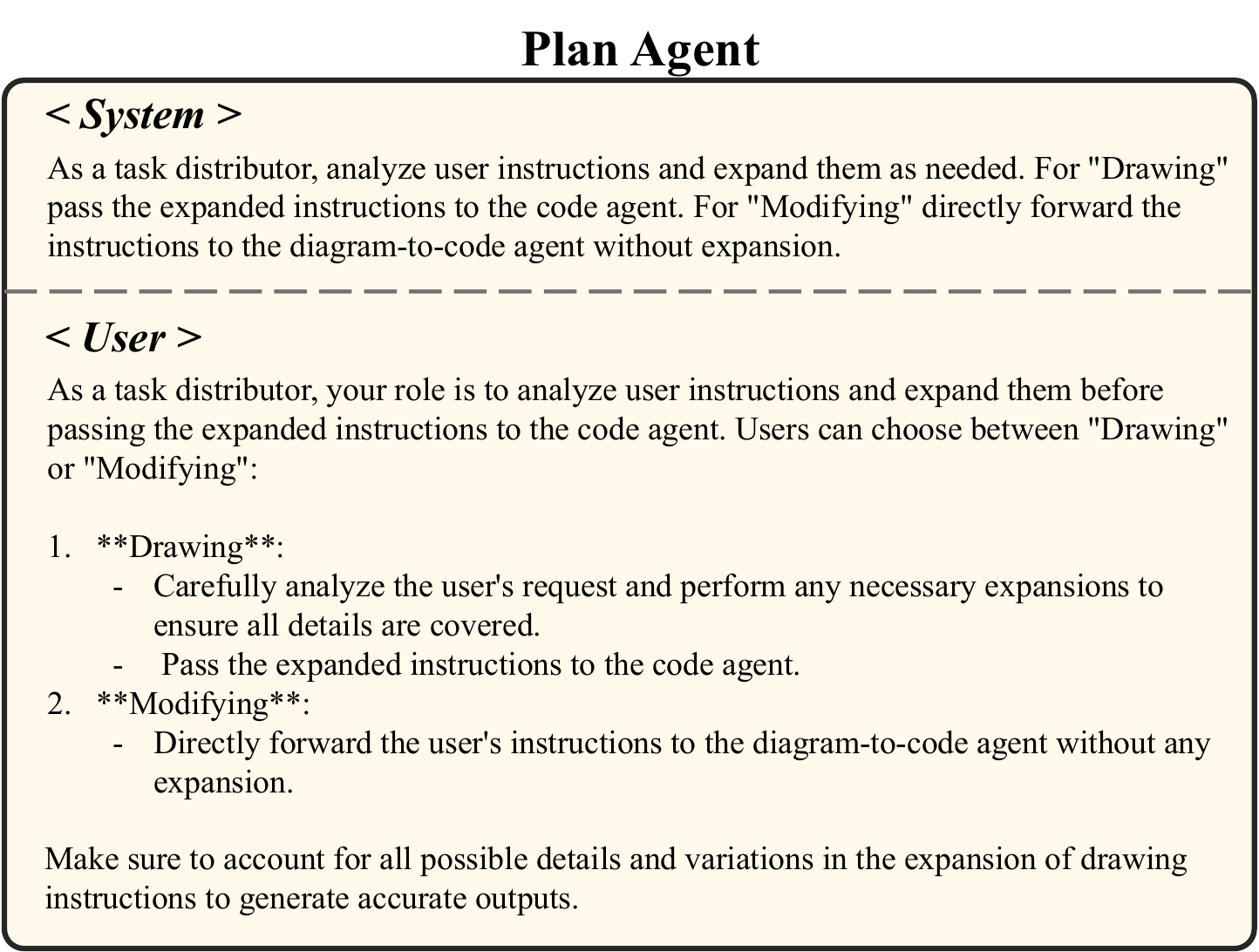}
    \caption{Prompt for Plan Agent: Analyzing and expanding user instructions based on the type of task (Generation or Editing) to ensure clarity before passing to the code generation stage.}
    \label{fig:plan_agent_prompt}
\end{figure}

\begin{figure}[h]
    \centering
    \includegraphics[width=\linewidth]{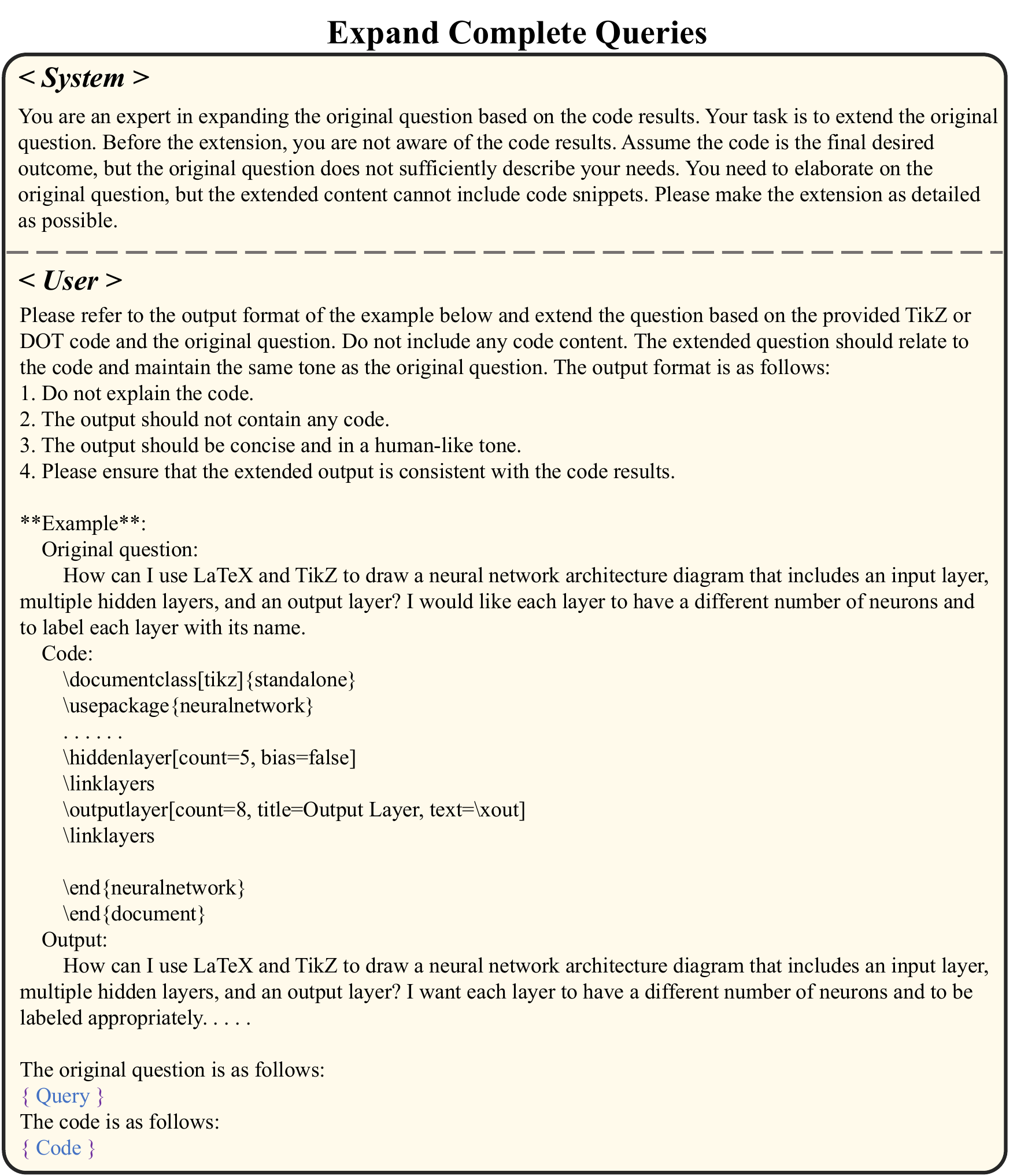}
    \caption{Prompt for Expanding Drawing Query to Generate Complete Query.}
    \label{fig:complete-query-prompt}
\end{figure}

\begin{figure}[ht]
    \centering
    \includegraphics[width=\linewidth]{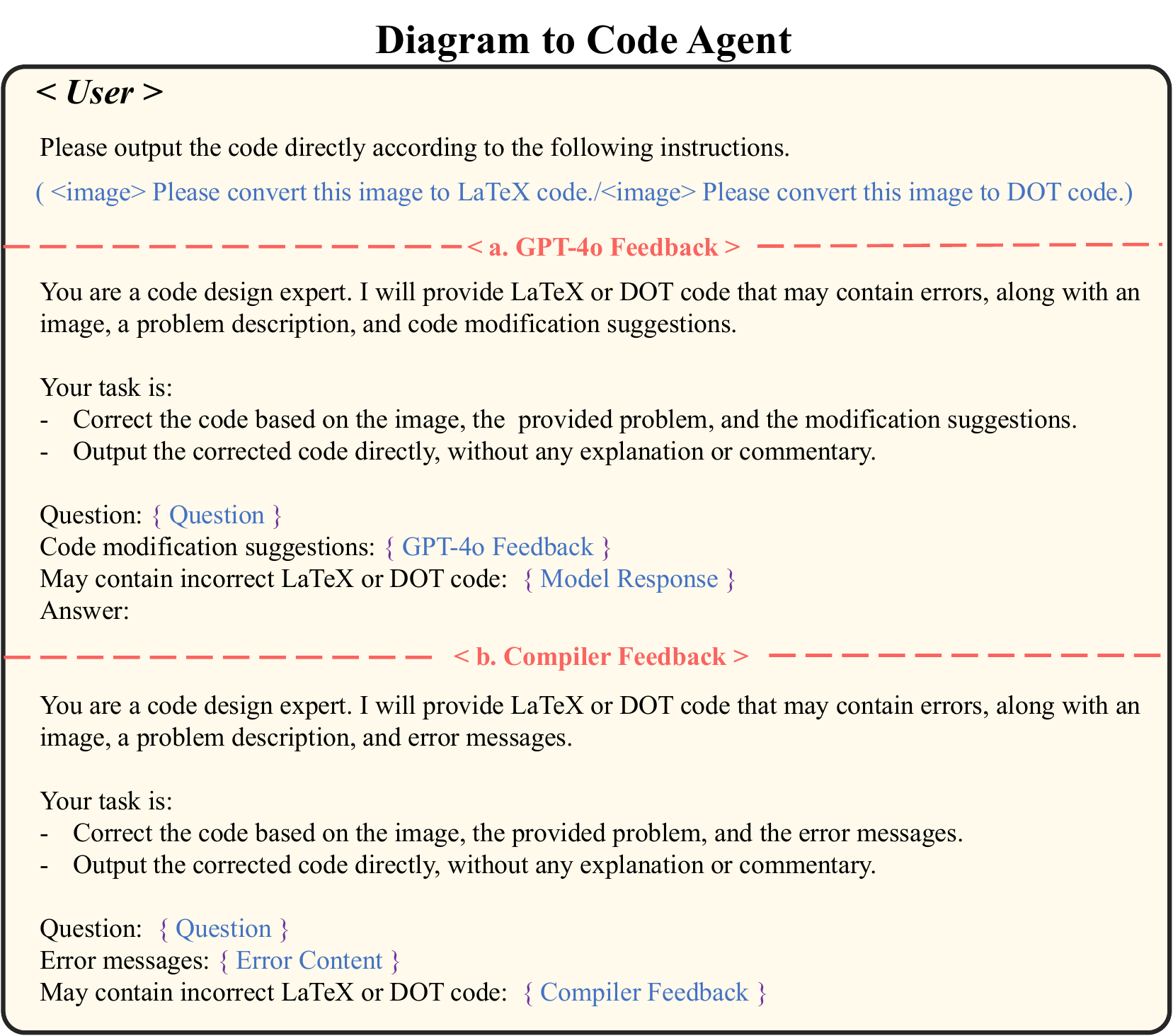}
    \caption{Prompt for Diagram-to-Code Agent: Handling direct conversion of diagrams to LaTeX or DOT code based on user-provided images and instructions.}
    \label{fig:diagram_to_code_agent_prompt}
\end{figure}

\begin{figure}[ht]
    \centering
    \includegraphics[width=\linewidth]{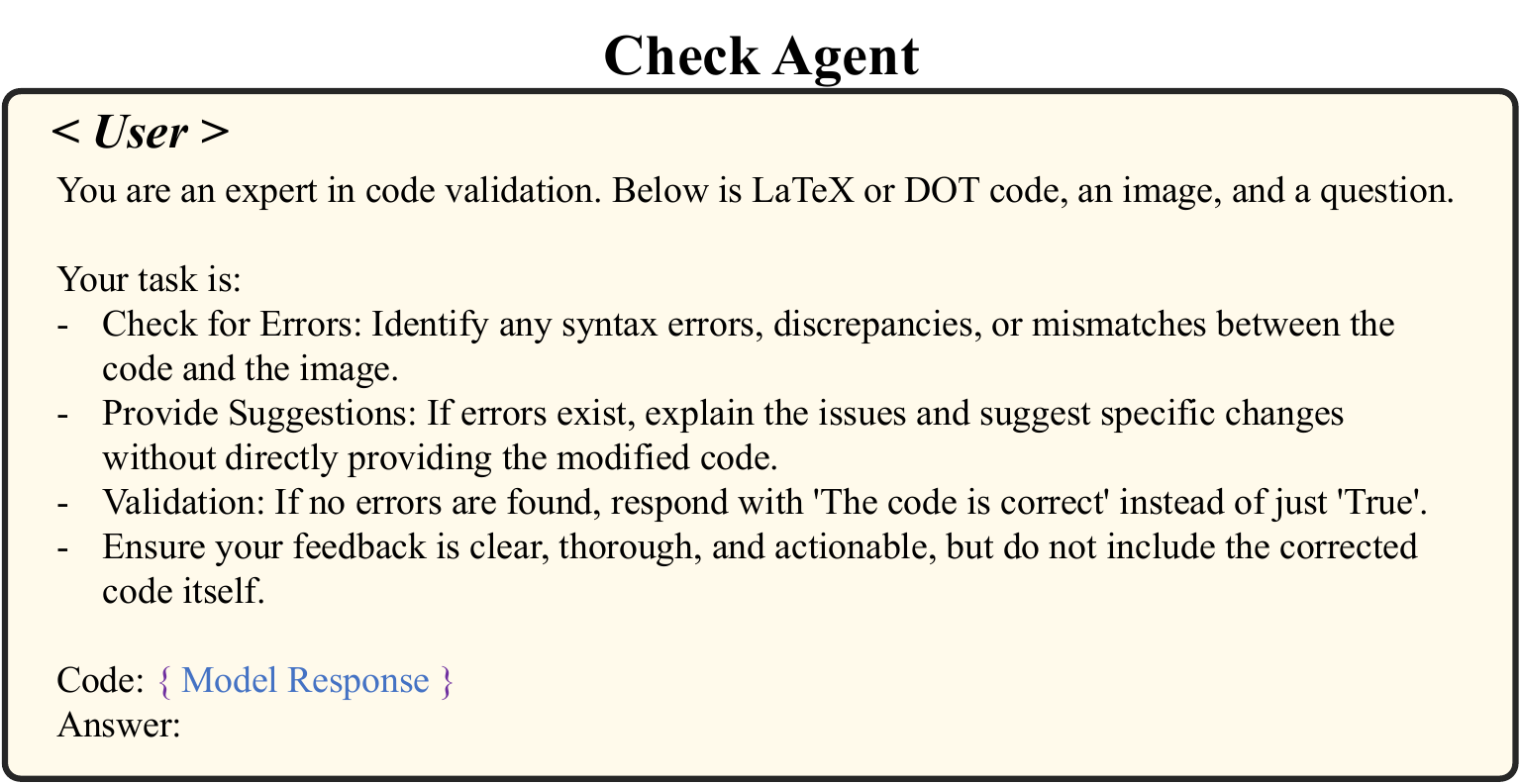}
    \caption{Check Agent Prompt for Diagram Coding: Verifying code accuracy for Diagram-to-Code Agent outputs, ensuring logical consistency and correctness.}
    \label{fig:diagram_to_code_check_agent}
\end{figure}

\begin{figure}[ht]
    \centering
    \includegraphics[width=\linewidth]{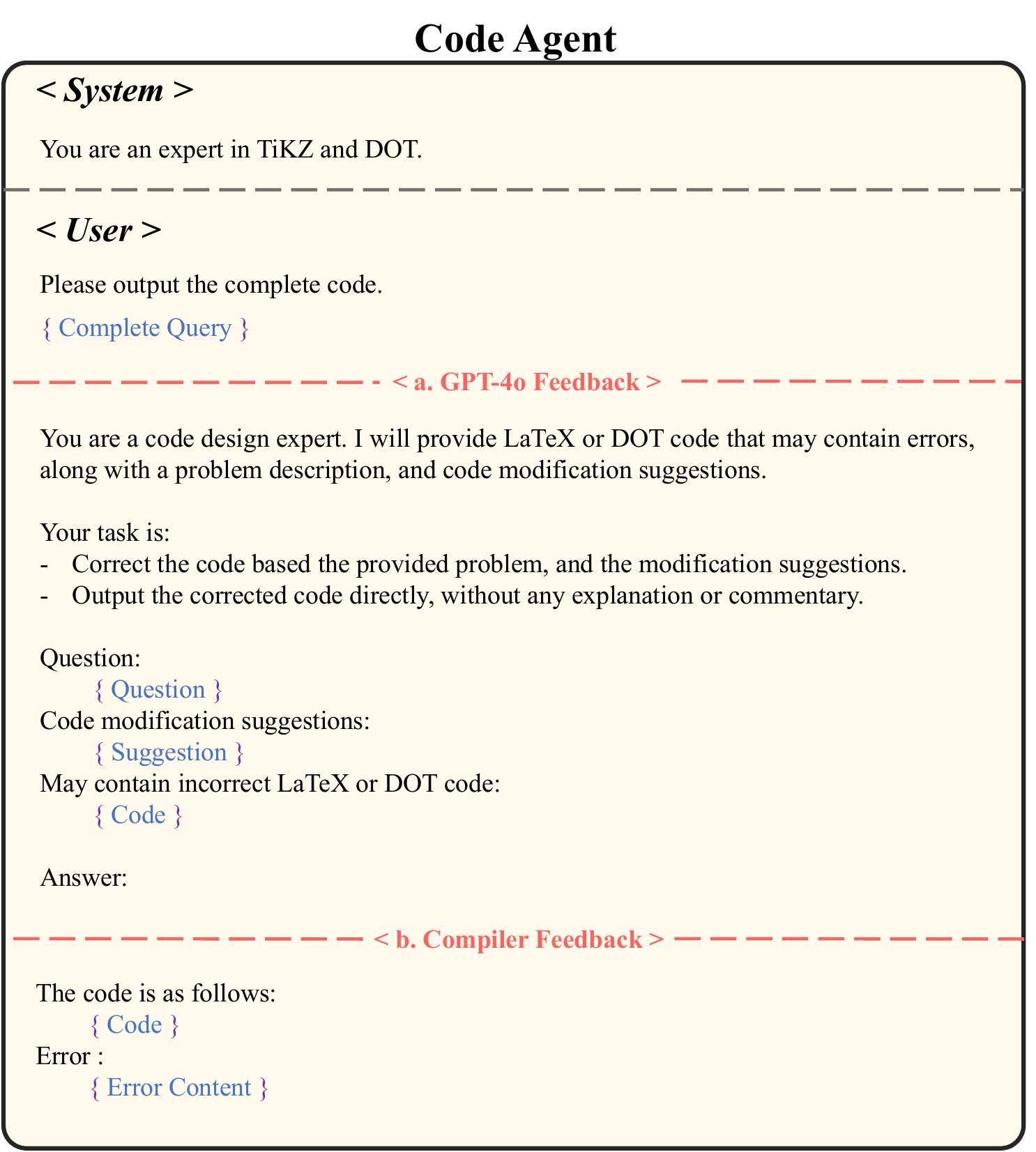}
    \caption{Code Agent Prompt for Diagram Generation: Generating code based on detailed instructions provided by the user.}
    \label{fig:instruction_to_code_agent}
\end{figure}

\begin{figure}[ht]
    \centering
    \includegraphics[width=\linewidth]{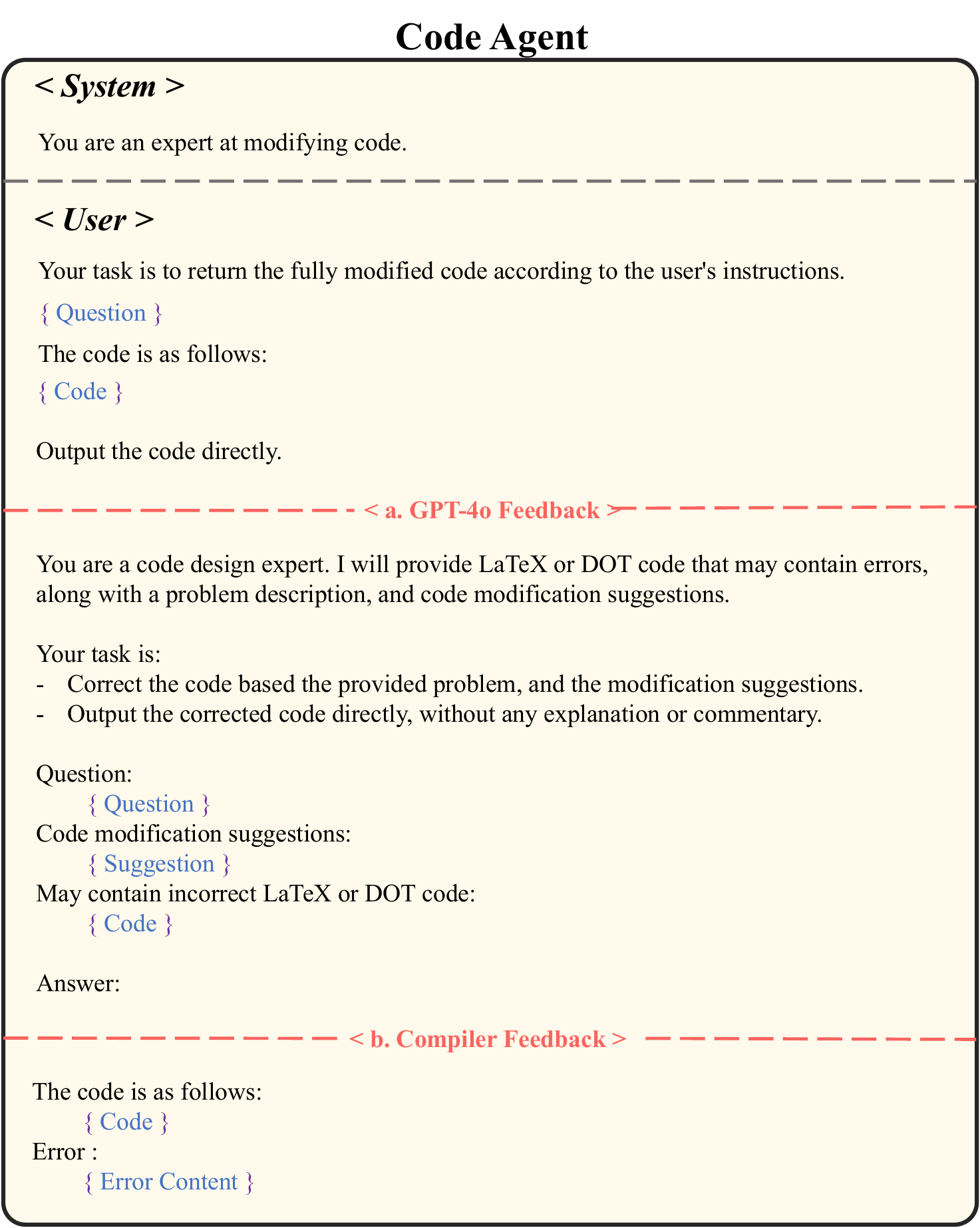}
    \caption{Code Agent Prompt for Diagram Editing: Handling feedback-based adjustments to existing code by applying suggestions from either GPT-4o or compiler feedback.}
    \label{fig:code_modification_task_prompt}
\end{figure}

\begin{figure}[ht]
    \centering
    \includegraphics[width=\linewidth]{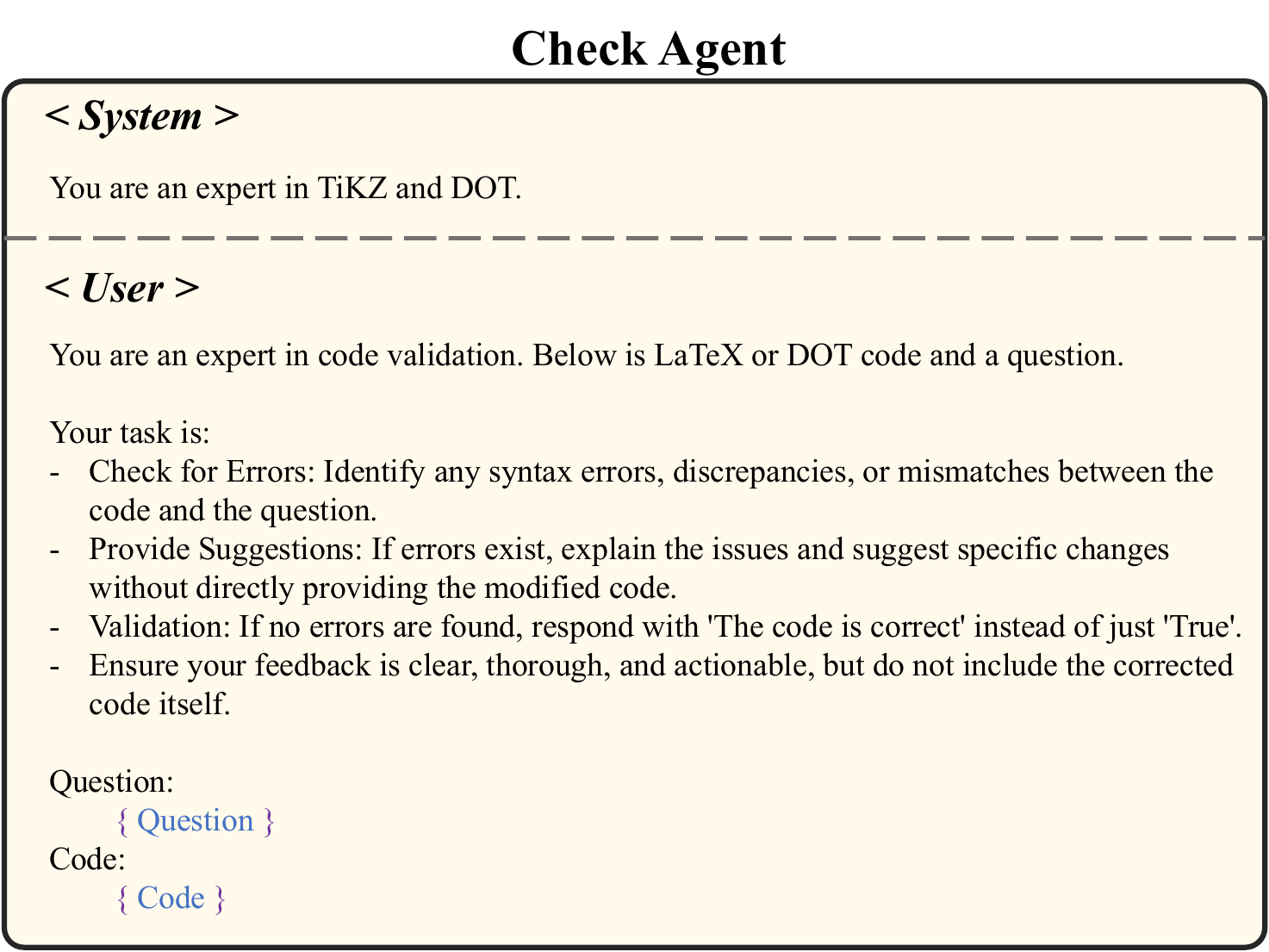}
    \caption{Check Agent: Using GPT-4o to perform error detection and provide correction suggestions within Code Agent outputs.}
    \label{fig:code_agent_check_agent}
\end{figure}

\begin{table*}[ht]
    \footnotesize
    \centering
    \caption{Comparison of DiagramGenBenchmark with Existing Benchmarks. "NL" indicates Natural Language inputs, "I" represents Image inputs, and "Code" denotes code output. Evaluation metrics vary across benchmarks, including pass rate, multi-level metrics, and other task-specific measures.}
    \label{tab:comparison}
    \setlength{\tabcolsep}{0.4mm}{
    \begin{tabular}{lcccccc}
        \toprule
        \textbf{Benchmark} & \textbf{Source} & \textbf{\# of Chart Types} & \textbf{\# of Test Instances} & \textbf{Input Format} & \textbf{Output Format} & \textbf{Evaluation Metric} \\
        \midrule
        HumanEval~\cite{chen2021evaluating} & Human Curated & - & 164 & Code & Code & Pass Rate \\
        MBPP~\cite{austin2021program} & Human Curated & - & 500 & NL+Code & Code & Pass Rate \\
        MMCode~\cite{li2024mmcode} & Crawl & - & 263 & NL+Code & Code & Pass Rate \\
        MatPlotBench~\cite{yang2024matplotagent} & Human Curated & 13 & 100 & NL & Code & GPT-4V Score \\
        Design2Code~\cite{si2024design2code} & Crawl & - & 484 & I+NL & Code & Multi-Level \\
        Plot2Code~\cite{wu2024plot2code} & Human Curated & 6 & 132 & I & Code & Pass Rate, Text-Match, etc. \\
        \midrule
        Diagram generation (Ours) & Human Curated & 8 & 270 & NL & Code,Diagram & Pass Rate, CodeBLEU, ROUGE-L, etc. \\
        Diagram coding (Ours) & Human Curated & 8 & 270 & I & Code & Pass Rate, CodeBLEU, ROUGE-L, etc. \\
        Diagram eidting (Ours) & Human Curated & 8 & 200 & NL+Code & Code,Diagram & Pass Rate, CodeBLEU, ROUGE-L, etc. \\
        \bottomrule
    \end{tabular}}
\end{table*}

\section{Comparison with Existing Benchmarks}

As shown in Table~\ref{tab:comparison}, DiagramGenBenchmark differs from existing benchmarks by providing both natural language and image inputs specifically for structured diagram generation and editing. Unlike traditional code generation benchmarks focused on simple text-based tasks, our benchmark encompasses a diverse range of diagram types and incorporates multi-level evaluation metrics (e.g., Pass@1, CodeBLEU, CLIP-FID, PSNR). Additionally, DiagramGenBenchmark includes three specialized tasks—diagram generation, diagram coding, and diagram editing—designed to evaluate both diagram creation and editing capabilities, making it a comprehensive resource for advancing multimodal diagram generation research.

% \newpage
% \clearpage
\section{Human Evaluation Details}

To evaluate the performance of DiagramAgent, three evaluators with master’s degrees independently rated the quality of generated diagrams on a scale from 1 to 5. The evaluation focused on the similarity between the generated diagrams and reference diagrams for two core tasks: the diagram generation and the diagram editing. Each score represents the following specific criteria:

\begin{itemize}
    \item \textbf{1}: The generated diagram shows very low similarity to the reference diagram, containing significant logical or structural errors and failing to fulfill the intended instructions.
    \item \textbf{2}: The generated diagram partially reflects the intended design but exhibits several issues, including low accuracy and visual coherence, and falls below the expected quality standards.
    \item \textbf{3}: The generated diagram meets basic requirements for structure and intent, though it has noticeable inconsistencies or errors that reduce its overall quality.
    \item \textbf{4}: The generated diagram is mostly accurate and visually coherent, with only minor, negligible errors; it aligns closely with the intended reference.
    \item \textbf{5}: The generated diagram is completely accurate and highly similar to the reference diagram, with no discernible errors; it fully meets or exceeds expectations for clarity and precision.
\end{itemize}

Each evaluator independently scored each of the $N$ test items, resulting in three scores per item. To calculate the mean score for each item, we average the three scores from the evaluators:

\begin{equation}
\text{Mean Score for Item } i = \frac{S^{(i)}_{1} + S^{(i)}_{2} + S^{(i)}_{3}}{3}
\label{eq:item_mean_score}
\end{equation}

where $S^{(i)}_{1}$, $S^{(i)}_{2}$, and $S^{(i)}_{3}$ represent the scores given by the three evaluators for item $i$.

After obtaining the mean score for each item, we calculate the overall average score for the model across all $N$ test items. This final score is computed as:

\begin{equation}
\text{Final Score for Model} = \frac{\sum_{i=1}^{N} \text{Mean Score for Item } i}{N}
\label{eq:model_final_score}
\end{equation}

This final score provides a standardized measure of the model’s ability to generate diagrams that closely resemble the reference diagrams. It reflects the model's performance across each task.

\newpage

\section{Diagram Coding Complete Results}
\label{app:image-to-code-complete}

This appendix provides a complete analysis of the experimental results for the diagram coding task with DiagramAgent, expanding on the brief introduction provided in Section \cref{Diagram_Coding} of the main text. Due to space constraints, only key findings were highlighted in the main text, while this appendix presents a more in-depth evaluation of DiagramAgent’s performance relative to other models across various metrics, as shown in Tables \ref{tab:image-to-code-main-complete} and \ref{tab:image-to-code-ablation-complete}.

\begin{table*}[h]
\footnotesize
\centering
\caption{Main results for diagram coding task (Diagram-to-Code Agent). The best result in each metric is bolded.}
\label{tab:image-to-code-main-complete}
% \resizebox{\linewidth}{!}{
\setlength{\tabcolsep}{4.2mm}{
\begin{tabular}{l|c|c|c|c|c|c|c}
\toprule
\textbf{Model} & \textbf{Size} & \textbf{Pass@1↑} & \textbf{ROUGE-L↑} & \textbf{codeBLEU↑} & \textbf{Edit Dist.↓} & \textbf{chrF↑} & \textbf{RUBY↑} \\
\midrule
Yi-VL~\cite{yivl} & 34B & 2.22 & 20.01 & 70.57 & 95.43 & 11.68 & 12.53 \\
Qwen2-VL~\cite{qwen2-vl} & 8B & 28.89 & 31.74 & 80.04 & 88.13 & 28.39 & 21.21 \\
Internlm-xcomposer2.5~\cite{zhang2024internlm} & 7B & 3.33 & 28.47 & 77.35 & 92.35 & 18.74 & 17.97 \\
Llama-3.2-Vision~\cite{llama3} & 11B & 27.78 & 21.94 & 75.37 & 92.92 & 16.37 & 13.95 \\
Phi-3.5-vision~\cite{phi3} & 4B & 24.07 & 27.53 & 76.56 & 90.01 & 20.86 & 17.96 \\
Llava-v1.6~\cite{liu2024llavanext} & 34B & 8.89 & 26.68 & 76.53 & 93.46 & 21.00 & 16.30 \\
Cogvlm2-llama3~\cite{hong2024cogvlm2} & 19B & 3.70 & 14.42 & 70.72 & 97.07 & 8.27 & 8.91 \\
Deepseek-vl~\cite{deepseekvl} & 7B & 50.74 & 25.18 & 76.48 & 88.82 & 18.35 & 16.13 \\
\midrule
GPT-4o~\cite{gpt4} & - & 64.07 & 39.95 & 81.78 & 86.68 & 34.40 & 26.18 \\
GLM-4-plus~\cite{chatglm} & - & 51.48 & 35.92 & 80.16 & 86.12 & 29.10 & 24.60 \\
Gemini-1.5-pro~\cite{gemini} & - & 17.78 & 38.66 & 80.75 & 88.05 & 30.00 & 25.62 \\
\midrule
\textbf{DiagramAgent} & 7B & \textbf{68.89} & \textbf{48.99} & \textbf{84.64} & \textbf{72.74} & \textbf{46.98} & \textbf{37.46} \\
\bottomrule
\end{tabular}}
\end{table*}

\begin{table*}[h]
\footnotesize
\centering
\caption{Ablation study for diagram coding task (Diagram-to-Code Agent). Each result shows the performance of DiagramAgent under various component configurations, with the decrease in each metric from the full model indicated in parentheses.}
\label{tab:image-to-code-ablation-complete}
% \resizebox{\linewidth}{!}{
\setlength{\tabcolsep}{4.2mm}{
\begin{tabular}{l|c|c|c|c|c|c|c}
\toprule
\textbf{Diagram Coding} & \textbf{Size} & \textbf{Pass@1↑} & \textbf{ROUGE-L↑} & \textbf{codeBLEU↑} & \textbf{Edit Dist.↓} & \textbf{chrF↑} & \textbf{RUBY↑} \\
\midrule
DiagramAgent & 7B & \textbf{68.89} & \textbf{48.99} & \textbf{84.64} & 72.74 & \textbf{46.98} & \textbf{37.46} \\
\midrule
-- w/o GPT-4o & 7B & 62.59 & 48.71 & 84.57 & \textbf{72.69} & 46.49 & 37.22 \\
& & (-6.30) & (-0.28) & (-0.07) & (-0.05) & (-0.49) & (-0.24) \\
-- w/o Compiler & 7B & 53.33 & 48.46 & 84.52 & 73.51 & 46.91 & 36.66 \\
& & (-15.56) & (-0.53) & (-0.12) & (+0.77) & (-0.07) & (-0.80) \\
-- w/o GPT-4o \& Compiler & 7B & 52.59 & 47.81 & 84.21 & 74.56 & 45.28 & 35.81 \\
& & (-16.30) & (-1.18) & (-0.43) & (+1.82) & (-1.70) & (-1.65) \\
\bottomrule
\end{tabular}}
\end{table*}

\paragraph{Main Results}
Table \ref{tab:image-to-code-main-complete} summarizes the main results for the diagram coding task, where DiagramAgent consistently outperforms other models across all key evaluation metrics, demonstrating its robustness and precision in translating diagrams into code. For Pass@1, DiagramAgent achieves 68.89, which is significantly higher than the scores of open-source models, such as Qwen2-VL (28.89) and Llama-3.2-Vision (27.78), as well as the scores of closed-source models such as GPT-4o (64.07) and GLM-4-plus (51.48). This high Pass@1 indicates that DiagramAgent performs well in generating correct code on the first attempt, reducing the need for iterative refinements and thus enhancing efficiency in the coding process.

The ROUGE-L score of 48.99 for DiagramAgent highlights its strength in preserving the structural and sequential accuracy of generated code, outperforming models such as GLM-4-plus (35.92) and GPT-4o (39.95). ROUGE-L is crucial in tasks that require alignment with the target’s structural elements, and DiagramAgent’s high score reflects its ability to capture complex syntactic patterns essential for correct code generation from diagrams. Additionally, DiagramAgent attains a codeBLEU score of 84.64, the highest among all models, which captures both syntactic and semantic alignment with the target code. Compared to other models, such as Qwen2-VL (80.04) and Gemini-1.5-pro (80.75), DiagramAgent demonstrates a marked improvement in generating both structurally and functionally accurate code, underscoring its capability to handle the nuanced requirements of diagram coding tasks. In terms of Edit Distance, DiagramAgent achieves a score of 72.74, which is the lowest among all models, indicating that fewer modifications are needed to align generated code with the target. Lower Edit Distance highlights DiagramAgent’s initial output accuracy, minimizing post-generation adjustments and improving overall workflow efficiency. This metric contrasts with higher Edit Distances observed in other models, such as Cogvlm2-llama3 (97.07) and Yi-VL (95.43), which indicate a greater need for corrections. Additionally, DiagramAgent’s chrF score of 46.98, higher than all compared models, reflects the fine-grained character-level alignment achieved by DiagramAgent, essential for high-quality code generation. Finally, DiagramAgent’s RUBY score of 37.46 is the highest across models, underscoring the model’s robustness and its capacity to generate code that aligns well with human coding standards, surpassing both close-source models, such as GPT-4o (26.18) and open-source options like Qwen2-VL (21.21). These comprehensive results collectively affirm DiagramAgent’s superior performance in the diagram coding task across all evaluation metrics.

\paragraph{Ablation Study}
Table \ref{tab:image-to-code-ablation-complete} presents the ablation study results, which evaluate the contributions of the GPT-4o verification and compiler debugging modules to DiagramAgent’s overall performance.
When the GPT-4o verification module is removed, DiagramAgent’s Pass@1 score drops by 6.30, while ROUGE-L decreases slightly by 0.28. These reductions suggest that the verification module enhances first-attempt accuracy, as indicated by the drop in Pass@1, and plays a role in preserving the structural fidelity of generated code, reflected in the ROUGE-L score. The minor reduction in codeBLEU by 0.07 further points to the verification component’s contribution to maintaining both syntactic and semantic consistency, while the minimal changes in Edit Distance (+0.05) and chrF (-0.49) indicate that this module has a moderate impact on fine-grained accuracy and error reduction. Removing the compiler debugging module has a more substantial effect, leading to a 15.56 decrease in Pass@1 and a 0.53 drop in ROUGE-L. These changes reflect the compiler’s crucial role in achieving syntactic accuracy, as it likely identifies and corrects structural inconsistencies early in the generation process. The Edit Distance increase of +0.77 suggests that the absence of debugging requires more extensive post-generation edits to achieve target alignment. The drops in chrF (-0.07) and RUBY (-0.80) underscore the compiler’s importance in ensuring character-level precision and maintaining the overall quality of the generated code. When both the GPT-4o verification and compiler debugging components are removed, DiagramAgent’s performance declines across all metrics, with a Pass@1 decrease of 16.30 and an Edit Distance increase of +1.82. The additional decreases in ROUGE-L (-1.18), codeBLEU (-0.43), chrF (-1.70), and RUBY (-1.65) further underscore the importance of these components for achieving DiagramAgent’s high standards in code generation. Without these modules, the model’s outputs show a higher propensity for errors, requiring greater manual adjustment and impacting both structural and semantic accuracy. This configuration demonstrates the complementary roles of verification and debugging in maintaining DiagramAgent’s precision and robustness in diagram coding tasks.

% \clearpage
\section{Error Analysis}
\label{app:error-analysis}

\begin{figure*}[htbp]
    \centering
    \includegraphics[width=\textwidth]{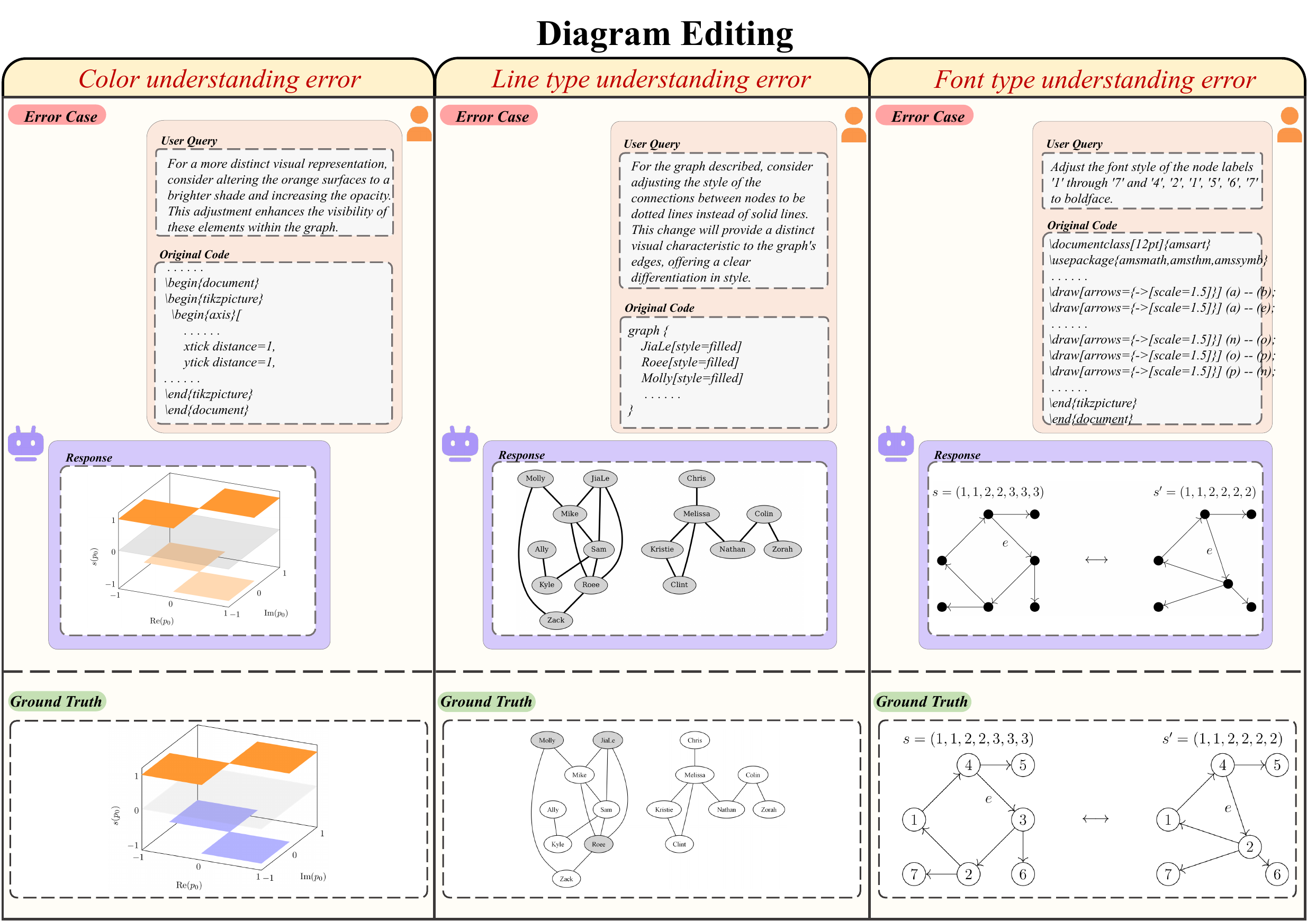}
    \caption{Common Error Types in the Diagram Editing: (Left) Color Understanding Error, (Center) Line Type Understanding Error, (Right) Font Type Understanding Error}
    \label{fig:code-modification-errors}
\end{figure*}

Our analysis of errors covers three primary tasks: diagram generation, diagram editing, and diagram coding, with common error types illustrated in Figures \ref{fig:instruction-to-code-errors}, \ref{fig:diagram-to-code-errors}, and \ref{fig:code-modification-errors}. Identifying and addressing these errors is crucial for enhancing the overall accuracy of generated diagrams, and it represents an essential focus for future work.

\subsection{Diagram Generation Errors}

\begin{figure*}[htbp]
    \centering
    \begin{minipage}[t]{\textwidth}
        \centering
        \includegraphics[width=\textwidth]{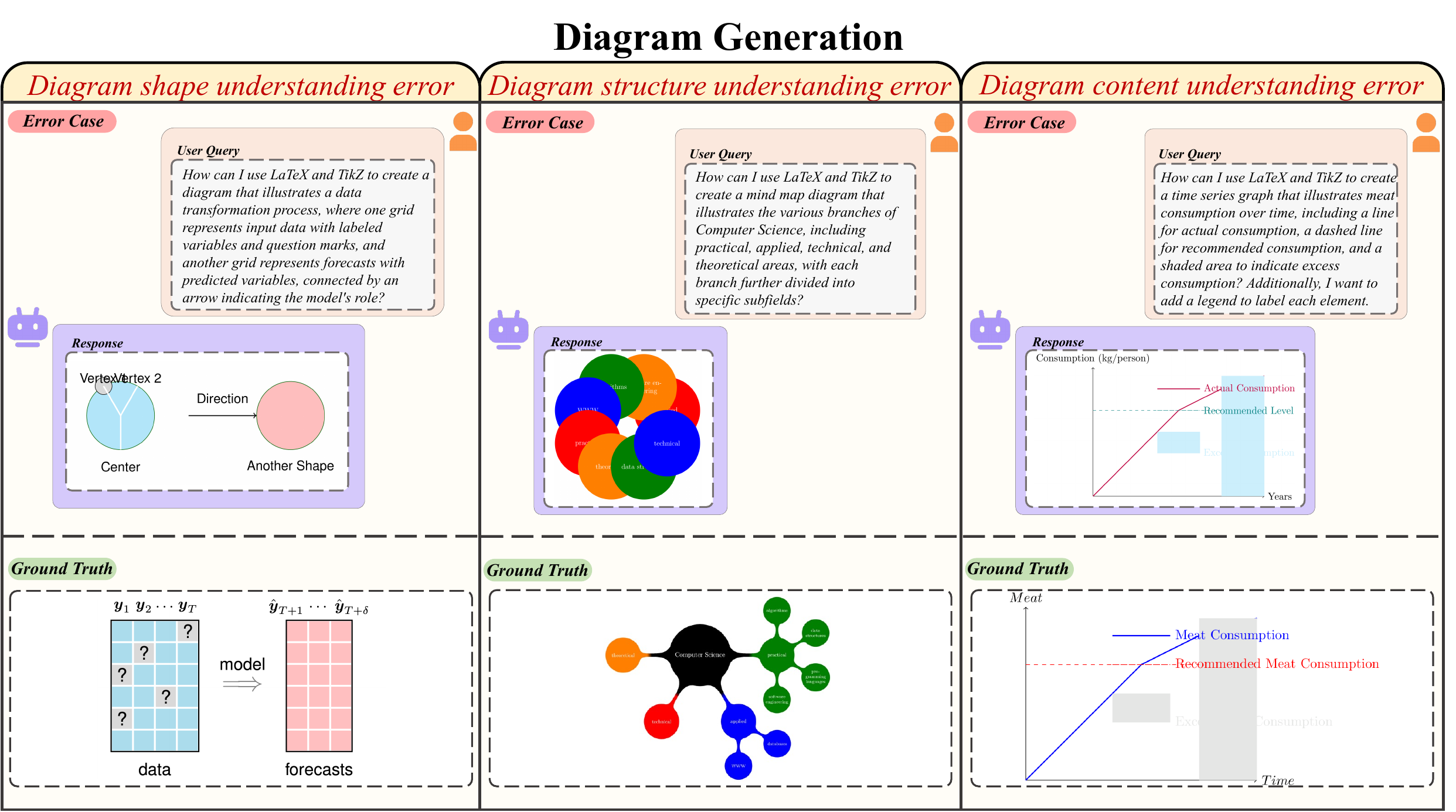}
        \caption{Common Error Types in the Diagram Generation: (Left) Shape Understanding Error, (Center) Structure Understanding Error, (Right) Content Understanding Error}
        \label{fig:instruction-to-code-errors}
    \end{minipage}
    % \qquad
    %让图片换行，
    \begin{minipage}[t]{\textwidth}
        \centering
        \includegraphics[width=\textwidth]{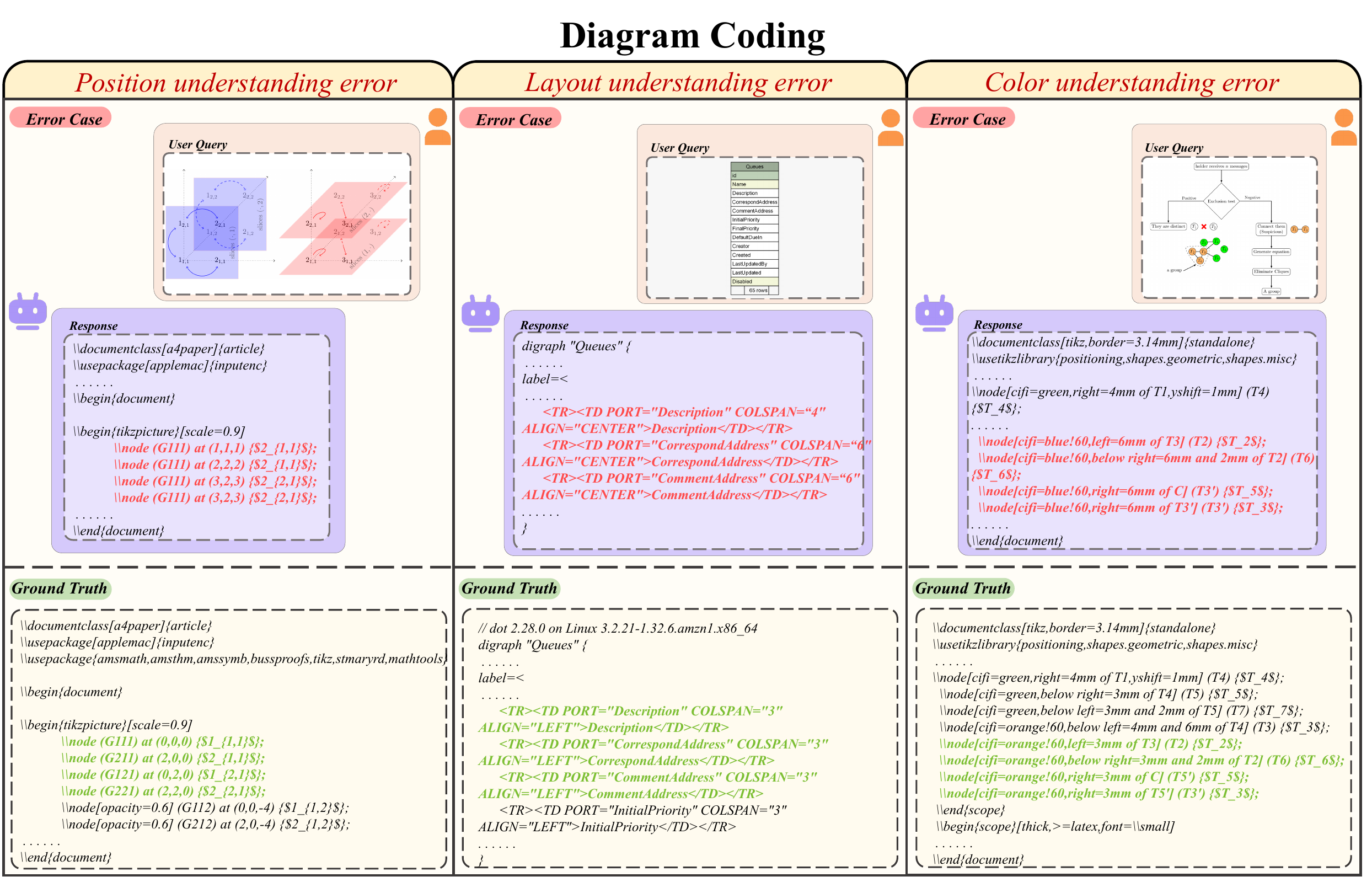}
        \caption{Common Error Types in the Diagram Coding: (Left) Position Understanding Error, (Center) Layout Understanding Error, (Right) Color Understanding Error}
        \label{fig:diagram-to-code-errors}
    \end{minipage}
\end{figure*}

For the diagram generation, three prominent error types emerge:
\begin{itemize}
    \item \textbf{Diagram Shape Understanding Error}: This error occurs when the model misinterprets the shape requirements specified in the user query. For instance, the model may generate incorrect or overly simplified shapes, as seen in Figure \ref{fig:instruction-to-code-errors} (left). This typically arises from ambiguity in the instruction or ineffective shape recognition capabilities.
    
    \item \textbf{Diagram Structure Understanding Error}: The model occasionally fails to capture the hierarchical or relational structure of diagrams, leading to incorrect element placements or connections, as shown in Figure \ref{fig:instruction-to-code-errors} (center). This error is often due to limited structural parsing of complex diagram elements. 
    
    \item \textbf{Diagram Content Understanding Error}: Errors in content understanding arise when the model misinterprets content-specific details, such as labels or numeric data, leading to inaccuracies in representation, as illustrated in Figure \ref{fig:instruction-to-code-errors} (right). This may result from insufficient parsing of content semantics in instructions. 
\end{itemize}

\subsection{Diagram Coding Errors}

For the diagram coding, errors typically involve spatial and layout issues:
\begin{itemize}
    \item \textbf{Position Understanding Error}: Misinterpretations of element positions result in misplaced components, as shown in Figure \ref{fig:diagram-to-code-errors} (left). This error often stems from insufficient spatial understanding of diagram elements. 
    
    \item \textbf{Layout Understanding Error}: Incorrect layout arrangements, where elements are poorly aligned or spaced, disrupt diagram coherence, as illustrated in Figure \ref{fig:diagram-to-code-errors} (center). This error arises from limitations in layout parsing and spatial arrangement. 
    
    \item \textbf{Color Understanding Error}: Similar to the diagram editing, color errors also appear in diagram coding, leading to misinterpreted color schemes that impact semantic meaning, as seen in Figure \ref{fig:diagram-to-code-errors} (right). This is due to inadequate interpretation of color attributes in diagrams. 
\end{itemize}

\subsection{Diagram Editing Errors}

In the diagram editing, the following errors are most prevalent:
\begin{itemize}
    \item \textbf{Color Understanding Error}: Color misinterpretation often leads to incorrect assignments, reducing visual clarity, as shown in Figure \ref{fig:code-modification-errors} (left). This issue typically arises from inadequate parsing of color-related instructions. 
    
    \item \textbf{Line Type Understanding Error}: Misunderstandings in line type specifications, such as solid vs. dashed lines, lead to errors in edge representation, as seen in Figure \ref{fig:code-modification-errors} (center). This stems from insufficient differentiation between line types in code interpretation. 
    
    \item \textbf{Font Type Understanding Error}: Incorrect font usage in diagrams, especially for labels, affects readability and visual coherence, as illustrated in Figure \ref{fig:code-modification-errors} (right). This issue is often due to limitations in font parsing and application. 
\end{itemize}

Addressing these errors is essential for enhancing the precision and fidelity of the generated diagrams. Future work should focus on refining the model’s capabilities in shape, structure, content, color, line type, font, position, and layout understanding to better align generated diagrams with user expectations.

% \begin{figure*}[pb]
%     \centering
%     \includegraphics[width=\textwidth]{img_appendix/img9.pdf}
%     \caption{Common Error Types in the Diagram Editing: (Left) Color Understanding Error, (Center) Line Type Understanding Error, (Right) Font Type Understanding Error}
%     \label{fig:code-modification-errors}
% \end{figure*}

% \begin{figure*}[htbp]
%     \centering
%     \begin{minipage}[t]{\textwidth}
%         \centering
%         \includegraphics[width=\textwidth]{img_appendix/img8.pdf}
%         \caption{Common Error Types in the Diagram Generation: (Left) Shape Understanding Error, (Center) Structure Understanding Error, (Right) Content Understanding Error}
%         \label{fig:instruction-to-code-errors}
%     \end{minipage}
%     %\qquad
%     %让图片换行，
%     \begin{minipage}[t]{\textwidth}
%         \centering
%         \includegraphics[width=\textwidth]{img_appendix/img10.pdf}
%         \caption{Common Error Types in the Diagram Coding: (Left) Position Understanding Error, (Center) Layout Understanding Error, (Right) Color Understanding Error}
%         \label{fig:diagram-to-code-errors}
%     \end{minipage}
% \end{figure*}